\documentclass[12pt,onecolumn,draftcls]{IEEEtran}
\usepackage[mathscr]{eucal}
\usepackage{ifpdf}
\usepackage{cite}
\usepackage{graphicx}
\usepackage[cmex10]{amsmath}
\usepackage{amssymb}
\usepackage{algorithmic}
\usepackage{units}
\usepackage{setspace}
\usepackage{algorithm}
\usepackage{array}
\usepackage[tight,footnotesize]{subfigure}
\usepackage{amsthm}
\usepackage{slashbox}
\usepackage{multirow}
\usepackage{enumerate}
\usepackage{color}

\newcounter{assume}

\newtheorem{theorem}{Theorem}
\newtheorem{lemma}{Lemma}
\newtheorem{proposition}{Proposition}
\newtheorem{corollary}{Corollary}

\newtheorem{assumption}[assume]{Assumption}
\newtheorem{remark}{Remark}

\newtheorem{definition}{Definition}

\usepackage{cancel}
\usepackage{soul}

\begin{document}
\title{Hybrid Centralized-Distributed Resource Allocation for
  Device-to-Device Communication Underlaying Cellular Networks}
%
\author{ \IEEEauthorblockN{Setareh Maghsudi and S\l awomir
   Sta\'{n}czak, \textit{Senior Member, IEEE}\\}
  \thanks{A short version of this paper is accepted to appear 
  at the IEEE International Conference on Communications, 2014. The 
  work was supported by the German Research Foundation (DFG) under 
  grant STA 864/3-3. The authors are with the Communications and 
  Information Theory Group, Technical University of Berlin. S. Sta\'{n}czak 
  is also with the Fraunhofer Institute for Telecommunications Heinrich 
  Hertz Institute, Berlin, Germany (e-mail: setareh.maghsudi@tu-berlin.de, 
  slawomir.stanczak@hhi.fraunhofer.de).}%
}
\maketitle
\begin{abstract}
The basic idea of device-to-device (D2D) communication is that pairs 
of suitably selected wireless devices reuse the cellular spectrum to 
establish direct communication links, provided that the adverse effects 
of D2D communication on cellular users is minimized and cellular users 
are given a higher priority in using limited wireless resources. Despite 
its great potential in terms of coverage and capacity performance, implementing 
this new concept poses some challenges, in particular with respect to 
radio resource management. The main challenges arise from a strong need 
for distributed D2D solutions that operate in the absence of precise 
channel and network knowledge. In order to address this challenge, this 
paper studies a resource allocation problem in a single-cell wireless 
network with multiple D2D users sharing the available radio frequency 
channels with cellular users. We consider a realistic scenario where 
the base station (BS) is provided with strictly limited channel knowledge 
while D2D and cellular users have no information. We prove a lower-bound 
for the cellular aggregate utility in the downlink with fixed BS power, 
which allows for decoupling the channel allocation and D2D power control 
problems. An efficient graph-theoretical approach is proposed to perform 
the channel allocation, which offers flexibility with respect to allocation 
criterion (aggregate utility maximization, fairness, quality of service 
guarantee). We model the power control problem as a multi-agent learning 
game. We show that the game is an exact potential game with noisy rewards, 
defined on a discrete strategy set, and characterize the set of Nash 
equilibria. Q-learning better-reply dynamics is then used to achieve equilibrium.
\end{abstract}
\section{Introduction}\label{sec:Introduction}
\subsection{Related Works}\label{subsec:RelatedWorks}
Device-to-device (D2D) communication as an underlay to cellular networks 
is regarded as one of the key technologies for enhancing the performance 
of future cellular networks \cite{fodor2012design}. The basic idea is to 
reuse cellular spectrum resources by allowing nearby wireless devices to 
establish direct communication links. This concept not only improves the 
efficiency of spectrum usage \cite{Kaufman08}, but also has a great potential 
for enhancing the network performance expressed in terms of capacity, coverage, 
energy efficiency and end-to-end delays \cite{Doppler09}. In 
order to realize networked-controlled D2D communication as an underlay to 
cellular networks, a system designer faces some challenges, which mainly 
arise due to the lack of reliable channel state information (CSI) at base 
stations (BS). In particular, efficient feedback is the key to obtaining 
CSI; nonetheless, while CSI for cellular users\footnote{In this paper, D2D 
user/link is used to refer to any pair of wireless devices that communicate 
directly, while any wireless device that operates in the traditional cellular 
mode is called a cellular user.}~can be efficiently acquired at a serving BS, 
such information is in general not available for D2D channels. The reason is 
the separation of the user/data plane from the control plane in the case of 
network-controlled D2D communication. An immediate consequence of this 
separation is that, in contrast to cellular users, D2D users cannot directly 
utilize pilot signals broadcasted by BSs for estimation of D2D channels. In 
addition, local transmissions of distinct pilot signals by each D2D user are 
infeasible and would not solve the problem due to pilot contamination.\footnote{Pilot 
contamination refers to a situation, in which the use of a large number of 
pilot signals leads to a relatively strong interference that may deteriorate 
the quality of channel estimation.}~Since strategies for suppressing pilot 
contamination in D2D scenarios suffer from the need for increased feedback 
and control overhead, it is reasonable to assume that allocation of resources 
to D2D users has to be performed in a distributed manner \textit{under strictly 
limited CSI}. Moreover, it is of utmost importance that direct transmissions 
among devices are coordinated to ensure that they do not have a detrimental 
impact on the performance of cellular users. Such coordination must involve a
careful power-controlled allocation of D2D users to available radio frequency 
channels, primarily used by a BS (downlink frequencies) and/or cellular users 
(uplink frequencies). This problem, which is difficult to solve even 
in a centralized manner, is further aggravated in D2D setting by the need for 
distributed solutions.

To date, numerous resource allocation schemes are developed for underlay D2D 
communication systems. Many of them, however, are only applicable to networks 
with limited number of D2D and/or cellular users. For instance, Reference \cite{Yu11} 
studies the optimal channel allocation and power control where one cellular and 
two D2D users share wireless resources. Similarly, in \cite{Xu12}, the system 
model includes one cellular and two D2D users, and a game-theoretical approach 
(reverse auction) is proposed to solve the resource sharing problem. References 
\cite{Belleschi11} and \cite{DFeng13} study a system with multiple D2D users; 
however, in every time slot, only one D2D user is allowed to transmit in a 
channel that is primarily allocated to a cellular user. Similar examples include 
\cite{Yu09}, \cite{Janis09} and \cite{Xu122}, among many others. 

Moreover, many works propose centralized resource allocation schemes for hybrid 
D2D and cellular communication. The schemes are mainly developed under the assumption 
that a central controller has access to the global channel and network knowledge, 
and therefore is capable of making coordination and resource allocation decisions 
not only for cellular users, but also for D2D users. For instance, Reference \cite{Phunchongharn13}, 
formulates the joint channel allocation and power control problem as a mixed integer 
programming, which is solved using column generation method. Similarly, in \cite{Aijaz14}, 
an energy-efficient uplink resource allocation scheme is proposed and analyzed by using 
mixed integer programming. The authors of \cite{Wang11} assume that a BS is able to 
perfectly coordinate the interference among cellular and D2D users. As 
another example, Reference \cite{Chen13} formulates a joint density and power allocation 
problem as a non-convex optimization problem using stochastic geometry, and proposes an 
algorithm to solve the problem. A joint resource allocation and mode selection mechanism 
based on particle swarm optimization is developed in \cite{Su13}. See also \cite{Feng13}, 
\cite{Han12} and \cite{Wang12} for further examples. 

In addition, in many research studies, some prior knowledge (such as information about 
utility functions) is assumed to be known to D2D users. In most cases, the problem is 
then solved using game-theoretical approaches such as pricing \cite{Wang13}, \cite{Wang14}, 
auctions \cite{Xu13} or coalition formation \cite{Song14}, \cite{Li14}, \cite{Chen14}, 
\cite{Cai14}. Moreover, Reference \cite{Huang14} proposes a resource allocation mechanism 
based on contract design. Besides requiring prior knowledge at the node level, most game-theoretical 
solutions impose large overhead due to the need for heavy information exchange in terms 
of bids, or prices and demands.

\subsection{Our Contribution}\label{subsec:Contribution}
The system model considered in this paper generalizes existing works in the 
following important directions:
\begin{itemize}
\item There is \textit{no limit} on the number of cellular and D2D users that 
coexist in the network.
\item \textit{Multiple} D2D users might be allowed to share a given channel with 
a cellular user.
\item The BS is only aware of \textit{statistical} channel knowledge of 
\textit{cellular users} and \textit{geographical locations} of D2D users. 
This information can be simply acquired by using pilot signals for cellular 
users and GPS (Global Positioning System) data of D2D users. 
This means that implementing D2D transmissions do not impose any overhead.
\item D2D and cellular users do not have \textit{any} channel knowledge.      
\end{itemize}

We first prove a lower-bound on the aggregate utility of cellular users. Based on 
this lower-bound, while taking the higher 
priority of cellular users into account, we decompose the resource allocation problem 
into two cascaded problems related to channel allocation and D2D power control. The 
former problem, which deals with maximizing the utility sum of cellular users, is a 
multi-objective combinatorial optimization problem that is very costly to solve with 
respect to the time and computational complexity. Therefore we propose a suboptimal, 
but efficient, graph-theoretical heuristic solution that involves maximum-weighted 
bipartite matching \cite{Kuhn55}, \cite{Galil86} and minimum-weighted graph partitioning 
\cite{Barnes'80}, \cite{Miller'91}. The problem can be then solved in a centralized 
manner by the BS, since the solution relies only on strictly limited information. The 
approach also offers high flexibility in terms of performance criteria, since quality 
of service or fairness can be also taken into account. The latter problem, in turn, 
deals with maximizing the aggregate utility of D2D users by means of power control, 
desirably in a distributed manner. We model the power control problem as a game with 
incomplete information, which, in contrast to most previous studies, is defined on 
a discrete strategy set. We show that this game is an exact potential game \cite{Ui08} 
and characterize the set of Nash equilibria. Furthermore, we use Q-learning better-reply 
dynamics \cite{Chapman13} in order to converge to Nash equilibrium. Finally, extensive 
numerical analysis is performed to evaluate the performance of the proposed approach 
in practical cases.

\subsection{Organization}\label{subsec:Organization}
The paper is organized as follows. In Section \ref{sec:SystemModel}, we 
introduce the network model and 
formulate the resource allocation problem. Section \ref{sec:Channel} 
is devoted to the first stage of the formulated problem, i.e., centralized 
channel allocation. Section \ref{sec:Power} deals with the second stage of 
the problem, i.e., distributed power control. Section \ref{sec:Numerical} 
presents numerical evaluations, while Section \ref{sec:conclusion} completes 
the paper.
\subsection{Notation}\label{subsec:Notation}
Throughout the paper we denote a set and its cardinality by a unique letter, 
and distinguish them by using calligraphic and italic fonts, such as $\mathcal{A}$ 
and $A$, respectively. Matrices are shown by bold upper case letters, for instance 
$\textbf{A}$. Moreover, $\textbf{A}_{l}$ denotes the $l$-th column of matrix 
$\textbf{A}$. Vectors are shown by bold lower case letters, for example $\textbf{a}$.

\section{System Model and Problem Formulation}\label{sec:SystemModel}
\subsection{System Model}\label{subsec:System}
\subsubsection{Network Model}\label{subsubsec:Network}
We consider the downlink of a \textit{single-cell} network with one BS denoted 
by $b$, and a set $\mathcal{L}$ consisting of $L$ single-antenna cellular users, 
each denoted by $l$. The cell is provided with a set $\mathcal{Q}$ of $Q=L$ 
orthogonal frequency channels that are referred to by $q$. Throughout 
the paper, by the term \textit{D2D user} we refer to a pre-defined pair of 
\textit{one single-antenna transmitter and one single-antenna receiver}, which 
is represented either by $k$ or by the pair $(k,k')$. Note that a single device 
can be either transmitter or receiver. We use $\mathcal{K}$ to denote the set 
of $K$ D2D users. The BS is able to communicate with multiple cellular users 
simultaneously, possibly by means of multiple antennas. The data stream intended 
to any given cellular user is transmitted with fixed average power 
$p_{c}$. Each D2D user selects a power level from the set $\mathcal{M}=
\left \{p_{d}^{(1)},p_{d}^{(2)},...,p_{d}^{(M)}\right \}$, where $1<p_{d}^{(1)}
<p_{d}^{(2)}<\cdots<p_{d}^{(M)}$. We assume that $p_{d}^{(M)}\ll p_{c}$, since 
in general the BS has access to larger energy resources in comparison with user 
devices. Each \textit{downlink} frequency channel $q$ is used i) by the BS in order to transmit to some set $\mathcal{L}_{q}\subseteq \mathcal{L}$ 
of $L_{q}$ cellular users, and ii) by a set $\mathcal{K}_{q}\subseteq \mathcal{K}$ 
of $K_{q}$ D2D users for direct communication. We assume that $L_{q}=1~\forall~q 
\in \mathcal{Q}$; that is, each channel is assigned to exactly one cellular user and 
therefore no vacant channel exists. This assumption is made in order to protect 
cellular users from an excessive interference due to a high BS power. We use 
$\mathbf{p}_{d,q}=(p_{1},...,p_{K_{q}})$ to denote the vector of transmit powers 
of the D2D users that transmit through channel $q$. Throughout the paper, $h_{uv,q}>0$ 
is the \textit{average gain} of channel $q$ from transmitter $u$ to receiver 
$v$. We assume that $h_{uv,q}=f_{uv,q}g_{uv}$, where $0<f_{uv,q} \leq 1$ and $0<g_{uv}
\leq 1$ stand for fast fading and path loss, respectively. We assume that the 
channel gains of any given link are drawn from a stationary distribution. Moreover, 
due to channel reciprocity, we have $h_{uv,q}=h_{vu,q}$. Signal-to-interference 
ratio (SIR) is denoted by $\gamma$. We consider a high SIR regime where $1<\gamma$, 
so that $\log\left (1+\gamma\right)\approx \log \left (\gamma \right)$. When treating 
interference as noise, $\log \left(\gamma \right)$ represents the achievable transmission 
rate of interference-limited point to point transmission.
\subsubsection{Utility Model}\label{subsubsec:Transmission}
The utility of cellular user $l \in \mathcal{L}_{q}$ that occupies channel 
$q$ is defined as\footnote{Throughout the paper, all logarithms are natural.}
\begin{equation}
\label{eq:CellThroughput}
R_{l}(q,\mathbf{p}_{d,q})=\log\left(\frac{p_{c}h_{bl,q}}{1+\sum_{k \in \mathcal{K}_{q}}
p_{k}h_{kl,q}}\right),
\end{equation}
which corresponds to the achievable transmission rate, as described 
before. Note that this utility model is widely used in litrature; see for example \cite{Chiang07}.
 
Since D2D users are subject to power control in addition 
to channel allocation, the utility of any D2D user $k \in \mathcal{K}_{q}$ 
is defined to be
\begin{equation}
\label{eq:DDThroughput}
R_{k}\left(q,\mathbf{p}_{d,q}\right)=\log \left(\frac{p_{k}h_{kk',q}}{1+\sum_{j 
\in \mathcal{K}_{q},j\neq k}p_{j}h_{jk',q}+p_{c}h_{bk',q}}\right)-c p_{k},    
\end{equation}
where $c$ is a fixed price factor to penalize excessive power usage \cite{Scutari06}. 
By definition, the utility of a D2D user corresponds to its transmission rate (see above) minus 
a cost that is paid to the cellular user in order to reimburse the adverse effects 
of spectrum sharing. The price factor can be either equal for all D2D users (as in 
(\ref{eq:DDThroughput})) or selected proportional to the channel gain (or distance) 
between a D2D user and the cellular user transmitting in the same channel \cite{Maghsudi12b}. 
Our analysis holds for both cases.
%
\subsubsection{Information Model}\label{subsubsec:Information}
We consider a model with strictly limited information, as described in the 
following assumption.
\begin{assumption}
\label{as:AvailInfo}
Each of the following is assumed throughout the paper.
\begin{enumerate}[a)] 
\item The BS has knowledge of i) geographical locations of cellular and D2D users 
and the path loss exponent, thereby $g_{lk}~\forall~l\in \mathcal{L},k \in \mathcal{K}$, 
and ii) the average fading gain of cellular to BS links, i.e., $h_{bl,q}~\forall~l \in 
\mathcal{L}, q \in \mathcal{Q}$.
\item The BS has no information about the fast fading component of cellular to 
cellular or D2D to D2D links.
\item Cellular and D2D users have no channel knowledge.
\end{enumerate}
\end{assumption}
\subsection{Problem Formulation}\label{subsec:Problem}
Network aggregate utility is conventionally regarded as a measure for evaluating 
the performance of resource management protocols \cite{Curescu08}, \cite{Rad08}, 
\cite{Ferragut14}, \cite{Cheung12}. Based on this criterion, the problem is to 
allocate channels and power levels to cellular and D2D users so as to maximize 
the network aggregate utility. With (\ref{eq:CellThroughput}) and (\ref{eq:DDThroughput}) 
in hand, this problem can be stated formally as 
\begin{equation}
\label{eq:CenProb}
\underset{\mathcal{L}_{q},\mathcal{K}_{q},\mathbf{p}_{d,q}}{\textup{maximize}}~~\sum_{q=1}^{Q}
\left (\sum_{l \in \mathcal{L}_{q}}R_{l}\left (q,\mathbf{p}_{d,q}\right)+ 
\sum_{k \in \mathcal{K}_{q}}R_{k}\left (q,\mathbf{p}_{d,q} \right)\right),
\end{equation}
where $\mathcal{L}_{q} \subseteq \mathcal{L}$, $\mathcal{K}_{q} \subseteq  \mathcal{K}$, 
$\mathbf{p}_{d,q} \in \bigotimes_{k=1}^{K_{q}} \left \{p_{d}^{(1)},...,p_{d}^{(M)}
\right\}$ and $\bigotimes$ denotes the Cartesian product. Note that unlike some previous 
works such as \cite{XuA12} and \cite{XuO13}, the utility functions defined here are 
\textit{user-specific}, i.e., the reward of any given channel differs to different users. 
As a result, the \textit{set} of D2D and cellular users allocated to each channel is 
required to be determined, and not just the \textit{number} of users. 

Such formulation however does not comply with the underlay D2D concept, and suffers 
from the following drawbacks that make it difficult or even impossible to deal with: 
i) The objective function in (\ref{eq:CenProb}) is not available at the BS due to the lack of 
information (see Assumption \ref{as:AvailInfo}), ii) The higher priority of cellular 
users is not taken into account, and iii) The objective function depends on both 
channel and power allocations that are mutually dependent. Therefore a solution to 
(\ref{eq:CenProb}) is difficult to obtain and is expected to be not amenable to 
distributed implementation. Our goal is therefore to develop a sophisticated heuristic 
approach. To this end, we first prove a lower-bound on the aggregate utility of cellular 
users that enables us to decouple the channel allocation and power control problems.
\begin{proposition}
\label{pr:lowerBound}
For any $\mathbf{p}_{d,q}, p_c$ and channel gains, we have
\begin{equation}
\label{eq:lowerBound}
\sum_{q=1}^{Q}\sum_{l \in \mathcal{L}_{q}}R_{l}(q,\mathbf{p}_{d,q})>\sum_{q=1}^{Q}\sum_{l \in \mathcal{L}_{q}} 
\log (p_{c}h_{bl,q})-\sum_{q=1}^{Q}\sum_{l \in \mathcal{L}_{q}}\sum_{k \in \mathcal{K}_{q}} p_{d}^{(M)}
g_{kl}.
\end{equation}
\end{proposition}
\begin{IEEEproof}
See Appendix \ref{sec:AppOne}.
\end{IEEEproof}
In words, the lower-bound in (\ref{eq:lowerBound}) corresponds to the worst-case 
scenario, in which all D2D users transmit at the maximum available power and the fast 
fading component of all D2D to cellular links equals one, thereby causing the maximum 
interference. Thus, for any realization of channel gains, the accuracy of the bound 
depends strongly on the range of the set of power levels $\mathcal{M}$, i.e., $p_{d}^{(M)}
-p_{d}^{(1)}$. Apart from this, as the bound does not depend on D2D power allocation 
and relies on the available information at the BS, it can serve as a basis for 
resource management. 

Since cellular users are assumed to have a higher priority and should 
be served first, we propose a two-step resource allocation strategy. In the first 
step, the objective is to maximize the lower-bound in (\ref{eq:lowerBound}) on the 
aggregate utility of cellular users. More precisely, given $p_{d}^{(M)}$, $p_c$ and 
imperfect channel knowledge, we aim at assigning channels to cellular and D2D users 
so as 
\begin{equation}
\label{eq:primaryGoal}
\underset{\mathcal{L}_{q},\mathcal{K}_{q}}{\textup{maximize}}~~~\sum_{q=1}^{Q}\sum_{l 
\in \mathcal{L}_{q}} \log \left (p_{c}h_{bl,q} \right )-\sum_{q=1}^{Q}\sum_{l \in 
\mathcal{L}_{q}}\sum_{k \in \mathcal{K}_{q}} p_{d}^{(M)}g_{kl},
\end{equation}
subject to
\begin{equation}
\label{eq:clustConOneT}
L_{q}=1, ~~ \forall~q \in \mathcal{Q}.
\end{equation}
This problem is investigated in Section \ref{sec:Channel}.

Once channels are allocated, in the second step we address the power control 
problem for D2D users, with the goal of maximizing the aggregate utility of D2D 
users as formalized below.
\begin{equation}
\label{eq:secondaryGoal}
\underset{\mathbf{p}_{d,q} \in \bigotimes_{k=1}^{K_{q}} \left \{p_{d}^{(1)},
...,p_{d}^{(M)}\right\}}{\textup{maximize}}~~\sum_{q=1}^{Q}\sum_{k \in \mathcal{K}_{q}}
R_{k}\left (q,\mathbf{p}_{d,q} \right).
\end{equation}
Section \ref{sec:Power} is devoted to this problem.

Summarizing, the resource allocation problem is decomposed into a channel allocation 
problem for all users followed by a power control problem for D2D users. As we see 
later, while the first problem is solved by the BS using a centralized method, the 
second problem is solved by D2D users in a distributed manner. Using such a two-stage 
scheme, not only a higher priority of cellular users is taken into account, but also 
D2D users utilize the assigned channels efficiently. Moreover, the limited available 
information is exploited with low computational effort.

\section{Channel Allocation}\label{sec:Channel}
This section deals with the first step of resource management, i.e., channel assignment 
with the goal of optimizing the performance of cellular users in terms of (\ref{eq:primaryGoal}). 
\subsection{The Channel Allocation Scheme}\label{subsec:ChannelAll} 
We notice that the first and second terms in (\ref{eq:primaryGoal}) are \textit{proportional} 
to the sum of the desired signals and interferences over all cellular users, respectively. 
Moreover, while the first term depends only on cellular users, the second term depends on 
D2D users as well. Roughly speaking, the problem in (\ref{eq:primaryGoal}) can be rephrased 
as $\underset{x,y}{\textup{maximize}}~~f(x)-g(x,y)$, where $x$ and $y$ respectively denote 
the cellular and D2D channel assignments. This problem is a multi-objective combinatorial 
optimization problem that is NP-hard and hence notoriously difficult to solve. Therefore 
we propose the following suboptimal, but simple and efficient, heuristic approach: 
At the beginning, we maximize the first term (weighted signal sum) so that the sets $\mathcal{L}_{q}$, 
$q\in \mathcal{Q}$, are defined. Afterwards, given $\mathcal{L}_{q}$, we allocate D2D users 
to frequency channels in a way that the second term (interference sum) is minimized. 
Formally,
\begin{equation}
\label{eq:First}
\underset{\mathcal{L}_{q}}{\textup{maximize}}~~~\sum_{q=1}^{Q}\sum_{l \in \mathcal{L}_{q}} 
\log \left (p_{c}h_{bl,q} \right)
\end{equation}
subject to (\ref{eq:clustConOneT}), and
\begin{equation}
\label{eq:Second}
\underset{\mathcal{K}_{q}}{\textup{minimize}}~~~\sum_{q=1}^{Q}\sum_{l \in \mathcal{L}_{q}}
\sum_{k \in \mathcal{K}_{q}}p_{d}^{(M)}g_{kl}.
\end{equation}

We call (\ref{eq:First}) and (\ref{eq:Second}) as \textit{assignment} and \textit{clustering} 
problems, respectively. In the next two subsections, we show that these problems 
boil down to two classic graph-theoretical problems on the induced network graph, 
namely \textit{maximum-weighted bipartite matching} and \textit{minimum-weighted 
partitioning}. 
\subsubsection{Assignment Problem}\label{subsec:assignment}
In the following, we show that problem (\ref{eq:First}) can be formulated as 
a weighted bipartite matching, defined below.
\begin{definition}[Weighted Bipartite Matching]
\label{de:WBM}
Let $G=(\mathcal{V},\mathcal{E})$ be a weighted bipartite graph where $\mathcal{V}= 
\mathcal{V}_{1}\cup \mathcal{V}_{2}$, $\mathcal{V}_{1} \cap \mathcal{V}_{2}=\varnothing$ 
and $\mathcal{E}\subseteq \mathcal{V}_{1} \times \mathcal{V}_{2}$. Each edge $e \in 
\mathcal{E}$ connecting any two vertices $x \in \mathcal{V}_{1}$ and $y \in \mathcal{V}_{2}$ 
is associated with some weight $w_{xy}$. The weights are gathered in the 
$V_{1} \times V_{2}$ graph matrix denoted by $\textbf{W}=[w_{xy}]$.\\
Matching: A matching is a subset $\mathcal{M} \subseteq \mathcal{E}$ such that $\forall 
v \in \mathcal{V}$ at most one edge in $\mathcal{M}$ is incident upon $v$.\\
Maximum Matching: A matching $\mathcal{M}$ such that every other matching $\mathcal{M}'$ 
satisfies $W_{\mathcal{M}'} \leq W_{\mathcal{M}}$, where $W_{\mathcal{M}}$ denotes the total weight 
of the selected edges for some matching $\mathcal{M}$.\\
Minimum Matching: A matching $\mathcal{M}$ such that every other matching $\mathcal{M}'$ 
satisfies $W_{\mathcal{M}}\leq W_{\mathcal{M}'}$. 
\end{definition}
%
Based on Definition \ref{de:WBM}, consider a bipartite graph $G_{L}(\mathcal{V},\mathcal{E})$, 
with $\mathcal{V}_{1}=\mathcal{L}$ (the set of cellular users) and $\mathcal{V}_{2}=\mathcal{Q}$ 
(the set of channels). The weight of the edge connecting $l \in \mathcal{L}$ and $q \in \mathcal{Q}$, 
$w_{lq}$, is defined as the weighted average gain of channel $q$ between the cellular user $l$ and 
the BS, i.e., $\log(p_{c}h_{bl,q})$. The problem is then to assign each cellular user a channel so that 
(\ref{eq:clustConOneT}) and (\ref{eq:First}) are satisfied. Let the assignment be presented by an 
$L \times Q$ assignment matrix $\mathbf{A}=\left [a_{lq}\right]$, where 
\begin{equation}
a_{lq}=\left\{\begin{matrix}
1 &  \textup{if}~l\in \mathcal{L}_{q}\\ 
0 & \textup{otherwise}
\end{matrix}\right..
\end{equation}
Therefore $\mathbf{A}$ satisfies the following constraints:
\begin{equation}
\label{eq:AsConOne}
\begin{matrix}
\sum_{l=1}^{L}a_{lq}\leq 1 & ,~q \in \left \{1,2,...,Q \right \}
\end{matrix},
\end{equation}
\begin{equation}
\label{eq:AsConTwo}
\begin{matrix}
\sum_{q=1}^{Q}a_{lq}= 1 & ,~l\in \left \{1,2,...,L \right \}
\end{matrix},
\end{equation}
\begin{equation}
\label{eq:AsConThree}
\begin{matrix}
a_{lq}\in \left \{0,1\right \} &,~\forall ~l,q
\end{matrix}.
\end{equation}
While (\ref{eq:AsConOne}) implies that each channel serves at most one cellular 
user, (\ref{eq:AsConTwo}) means that each cellular user is served by exactly 
one channel. Note that equality holds in (\ref{eq:AsConOne}) as we assume $Q=L$ 
(see Section \ref{subsubsec:Network}). The sum of edges' weights yields 
\begin{equation}
\label{eq:AsProb}
\sum_{q=1}^{Q}\sum_{l \in \mathcal{L}} w_{lq}a_{lq} = \sum_{q=1}^{Q}\sum_{l \in \mathcal{L}_{q}} w_{lq}.
\end{equation}
Thus, the problem in (\ref{eq:First}) subject to (\ref{eq:clustConOneT}) is equivalent 
to maximizing (\ref{eq:AsProb}), subject to (\ref{eq:AsConOne}), (\ref{eq:AsConTwo}), 
and (\ref{eq:AsConThree}), i.e., it corresponds to the maximum matching of $G_{L}$.
\subsubsection{Clustering Problem}\label{subsec:clustering}
This step consists of allocating channels to D2D users with the goal of minimizing 
the total interference to the cellular users over all channels. In order to address 
this problem we need to define the network graph.
\begin{definition}[Network Graph]
\label{de:PIG}
The network graph for any channel $q \in \mathcal{Q}$ is an undirected graph 
$G_{N}=(\mathcal{V},\mathcal{E})$ with $\mathcal{V}=\mathcal{V}_{1} \cup 
\mathcal{V}_{2}$, where $\mathcal{V}_{1}$ and $\mathcal{V}_{2}$ represent the 
set of $K$ D2D \textit{transmitters} and $L$ cellular receivers, respectively. 
The weight of an edge between any pair of graph vertices $(x,y)$ is denoted by 
$w_{xy}$, where $w_{xy}$ is equal to the average gain of channel $q$ between 
$x$ and $y$. 
\end{definition}
However, by Assumption \ref{as:AvailInfo}, only limited CSI is available at 
the BS; therefore the network graph cannot be constructed. As a result, we 
define the \textit{estimated} network graph, which can be reproduced by the 
BS using the available information.
\begin{definition}[Estimated Network Graph]
\label{de:EWIG}
Estimated network graph is an undirected graph $G_{E}=(\mathcal{V},\mathcal{E})$ 
with $\mathcal{V}=\mathcal{V}_{1} \cup \mathcal{V}_{2}$, where $\mathcal{V}_{1}$ 
and $\mathcal{V}_{2}$ represent the set of $K$ D2D transmitters and $L$ cellular 
receivers, respectively. The weight of an edge between any D2D transmitter $k$ and 
cellular receiver $l$ is defined as $w_{kl}=p_{d}^{(M)}g_{kl}$. The weight of the 
edge between any two cellular users and any two D2D users are respectively equal to 
some constant $C> Kp_{d}^{(M)}$ and zero.\footnote{Later we see that this definition 
results in some form of clustering by which the cellular to cellular and also the 
D2D to cellular interferences decrease. D2D to D2D interference is however neglected. 
This implies that in the absence of full and precise channel knowledge the priority 
is to protect cellular users.}
\end{definition}
Next we show that problem (\ref{eq:Second}) can be rephrased as Q-way 
minimum-weighted graph partitioning on the estimated network graph $G_{E}$.
\begin{definition}[Q-way Weighted Partitioning]
\label{de:LPP}
Let $G=(\mathcal{V},\mathcal{E})$ be a weighted graph where each edge $e 
\in \mathcal{E}$ connecting any two vertices $x$ and $y$ is associated with 
some weight $w_{xy}$. The weights are gathered in a $V \times V $ matrix denoted 
by $\mathbf{W}=[w_{xy}]$. The minimum-weighted Q-way partitioning problem 
divides the set of vertices into $Q$ disjoint subsets in a way that the sum 
weights of edges whose incident vertices fall into the same subset is minimized. 
\end{definition}
Now consider the estimated network graph, $G_{E}$. Then solving (\ref{eq:Second}) is 
equivalent to finding some $(L+K) \times Q$ assignment matrix $\mathbf{B}=\left 
[b_{jq}\right]$ that is defined to be
\begin{equation}
b_{jq}=\left\{\begin{matrix}
1 &  \textup{if} ~j \in \mathcal{L}_{q} \cup \mathcal{K}_{q} \\ 
0 & \textup{otherwise}
\end{matrix}\right..
\end{equation}
Thus each column in $\mathbf{B}$, e.g., $\mathbf{B}_{q}=\left [b_{1q},b_{2q},...,
b_{(L+K)q}\right]^{T},q\in \left \{ 1,2,...,Q\right \}$, is an indicator describing 
cluster $q$. Therefore $b_{jq}$ satisfies the following constraints:
\begin{equation}
\label{eq:clustConOne}
\begin{matrix}
\sum_{j=1}^{L+K}b_{jq}=L_{q}+K_{q}&,
~q\in \left \{1,2,...,Q\right\}
\end{matrix},
\end{equation}
\begin{equation}
\label{eq:clustConTwo}
\begin{matrix}
\sum_{q=1}^{Q}b_{jq}=1 ~~& ,~j\in \left \{1,2,...,L+K \right \}
\end{matrix},
\end{equation}
and
\begin{equation}
\label{eq:clustConThree}
\begin{matrix}
b_{jq}\in \left \{0,1 \right \}&,~\forall ~j,q.
\end{matrix}
\end{equation}
The sum of edges' weights connecting users in cluster $q$ hence follows as
\begin{equation}
\label{eq:clustObjectO}
\frac{1}{2}\sum_{j \in \mathcal{L} \cup \mathcal{K}}\sum_{j' \in \mathcal{L} \cup \mathcal{K}} 
w_{jj'}b_{jq}b_{j'q}=\frac{1}{2}\mathbf{B}_{q}^{T} \mathbf{W}_{E} \mathbf{B}_{q},
\end{equation}
where $\mathbf{W}_{E}$ is the weight matrix of $G_{E}$. As a result, the total 
sum-weight of edges that are not cut by the Q-way partitioning of $G_{E}$ yields 
\begin{equation}
\label{eq:clustObjectT}
\begin{aligned}
\frac{1}{2}\sum_{q=1}^{Q}\sum_{j \in \mathcal{L} \cup \mathcal{K}}\sum_{j' \in 
\mathcal{L} \cup \mathcal{K}} w_{jj'}b_{jq}b_{j'q}= &\\  
\frac{1}{2}\sum_{q=1}^{Q}\sum_{j \in \mathcal{K}}\sum_{j'\in \mathcal{K}} 
w_{jj'}b_{jq}b_{j'q}+\frac{1}{2}\sum_{q=1}^{Q}&\sum_{j \in \mathcal{L}}\sum_{j'\in \mathcal{L}}
w_{jj'}b_{jq}b_{j'q}\\ 
+2\times \frac{1}{2}\sum_{q=1}^{Q}\sum_{j \in \mathcal{L}}\sum_{j'\in \mathcal{K}}
w_{jj'}b_{jq}b_{j'q}.&
\end{aligned}
\end{equation}
The first term on the right-hand side of (\ref{eq:clustObjectT}) is zero by 
the definition of $G_{E}$. Also, by the following proposition, the second term 
equals zero as well, since any minimum-weighted partitioning assigns exactly 
one cellular user to each cluster. 
\begin{proposition}
\label{pr:nCellular}
Any minimum-weighted Q-way partitioning of the estimated network graph $G_{E}$ 
assigns exactly one cellular user to each cluster, that is $L_{q}=1~\forall 
q \in \mathcal{Q}$.
\end{proposition}
\begin{IEEEproof}
See Appendix \ref{sec:AppThree}.
\end{IEEEproof}
By Proposition \ref{pr:nCellular} and comparing (\ref{eq:clustObjectO}) with 
(\ref{eq:clustObjectT}) we have
\begin{equation}
\label{eq:clustObjectF}
\begin{aligned}
 \frac{1}{2}\sum_{q=1}^{Q}\mathbf{B}_{q}^{T} \mathbf{W}_{E} \mathbf{B}_{q}=
 \sum_{q=1}^{Q}\sum_{j \in \mathcal{L}}\sum_{j'\in \mathcal{K}}
&w_{jj'}b_{jq}b_{j'q} \\ 
 =\sum_{q=1}^{Q}\sum_{j \in \mathcal{L}_{q}}\sum_{j'\in \mathcal{K}_{q}}
&w_{jj'}. 
\end{aligned}
\end{equation}
By comparing (\ref{eq:clustObjectF}) with (\ref{eq:Second}) and by using the 
definition of $G_{E}$, it can be concluded that (\ref{eq:Second}) is equivalent 
to the minimum-weighted Q-way partitioning of $G_{E}$. 

\begin{remark}
As described in Section \ref{subsubsec:Network}, D2D user is referred to a 
\textit{pair} of one single-antenna transmitter and one single-antenna receiver. Also, as 
described before, after clustering, any transmitter-receiver pair, which represents a D2D 
user, belong to a single cluster. As a result, i) no D2D transmitter 
communicates simultaneously with multiple receivers, and ii) no inter-cluster 
communication takes place; that is, communication occurs only between devices in the same cluster.
\end{remark}
\subsection{Some Notes on Complexity}\label{subsec:Comp} 
In principal, the proposed channel allocation scheme solves two problems, namely maximum-weighted 
matching and minimum-weighted partitioning. The latter problem, however, can be itself reformulated 
as a minimum-weighted matching, due to the special characteristics of the defined estimated network 
graph. This is described formally in the following proposition. 
\begin{proposition}
\label{pr:conversion}
Define a bipartite graph $G'(\mathcal{V},\mathcal{E})$ where $\mathcal{V}_{1}=\mathcal{K}$ 
and $\mathcal{V}_{2}$ is produced by $K$ times replicating $\mathcal{L}$, i.e., $\mathcal{V}_{2}
=\underbrace{\mathcal{L} \cup \mathcal{L}...\cup \mathcal{L}}_{\times K}$. The weight of 
any edge connecting some D2D user $k \in \mathcal{V}_{1}$ to each copy $l_{j}\in \mathcal{V}_{2}$ 
($j \in \left \{1,...,K \right \}$) of some cellular user $l \in \mathcal{L}$ is $w_{lk}$, 
that is, equal to the weight of the edge connecting $k$ and $l$ in the estimated network graph, 
$G_{E}$. Then the minimum-weighted Q-way partitioning of $G_{E}$ is equivalent to a minimum-weighted 
bipartite matching of $G'$.
\end{proposition}
\begin{IEEEproof}
See Appendix \ref{sec:AppFour}.
\end{IEEEproof}
Therefore the algorithm is required to solve two (parallel) weighted matching problems. 
Weighted matching is a classic graph-theoretical problem for which numerous efficient 
algorithmic solutions exist. A well-known solution is the Hungarian algorithm \cite{Kuhn55}. 
For a bipartite graph $G(\mathcal{V},\mathcal{E})$, the space complexity of Hungarian 
algorithm yields $O(V^{2}E)$ with $V=\max\left \{V_{1},V_{2} \right\}$,\footnote{In case 
$V_{1}\neq V_{2}$, dummy vertices are added. See \cite{Kuhn55} for details.}~that is 
polynomial in the number of vertices and also in the number of edges. The running time 
is $O(V^3)$, which is also polynomial in the number of vertices. In our model, for 
the first matching we have $V=L$ and $E=L^{2}$, by the definition of $G_{L}$.\footnote{This 
number of edges corresponds to the worst-case scenario where the bipartite graph is complete, i.e., 
there exists an edge between any pair $x \in \mathcal{V}_{1}$ and $y \in \mathcal{V}_{2}$.}~For 
the second matching, on the other hand, we have $V=KL$ and $E=\left (KL\right )^{2}$, by 
the definition of $G_{E}$ and Proposition \ref{pr:conversion}. Note that the two problems 
can be solved simultaneously; hence the running times do not add up. More algorithmic 
solutions can be found in \cite{Galil86} and \cite{Micali80} for instance.
\subsection{Quality of Service Guarantee and Fairness}\label{subsec:Remarks} 
Despite being suboptimal, the decoupling approach described in Section \ref{subsec:ChannelAll} 
provides the possibility of solving the channel allocation problem efficiently under different 
constraints. Two examples are given below.
\begin{itemize}
\item Quality of service (QoS) requirement for cellular users: By problem (\ref{eq:primaryGoal}), 
the goal of channel allocation is to provide \textit{every} D2D user with some transmission 
channel in a way that the aggregate utility of cellular users is maximized, thereby ignoring 
the \textit{individual} performances of cellular users. In many networks, however, cellular 
users require some specific QoS that restricts the amount of tolerable interference. Assume 
that each cellular user $l$ requires some minimum utility, $R_{l,\textup{min}}$, by which its 
QoS is guaranteed. After solving problem (\ref{eq:First}), each cellular user is assigned a 
channel. Therefore, the nominator of (\ref{eq:CellThroughput}) is known. As a result, the 
maximum tolerable interference of each cellular user $l$, say $I_{l,\textup{max}}$, can be 
calculated based on $R_{l,\textup{min}}$. We construct a bipartite graph with $\mathcal{V}_{1}=\mathcal{K}$ 
and $\mathcal{V}_{2}=\mathcal{L}$. The problem is then to assign \textit{as many as possible} D2D users 
to cellular users (thus to channels) so that no interference experienced by any cellular 
user exceeds the maximum tolerable value. Formally, the problem is to find an $K \times L$ 
assignment matrix $\mathbf{X}=[x_{kl}]$ so that 
\begin{equation}
\textup{maximize}~~\sum_{l=1}^{L}\sum_{k=1}^{K} x_{kl},
\end{equation}
subject to the following constraints:
\begin{equation}
\sum_{k \in \mathcal{K}} w_{kl}~x_{kl} \leq I_{l,\textup{max}},~~ \forall l\in \mathcal{L},
\end{equation}
\begin{equation}
\sum_{l=1}^ {L} x_{kl} \leq 1,~~\forall k\in \mathcal{K},
\end{equation}
and 
\begin{equation}
x_{kl} \in \{0,1\},~~\forall l,k.
\end{equation}
Note that by the definition of estimated network graph, $w_{kl}=p_{d}^{(M)}g_{kl}$, 
i.e., it is an upper-bound of the interference experienced by cellular user $l$ due to 
D2D user $k$. This problem is known as the \textit{generalized assignment problem} 
which is NP-hard; nonetheless, efficient approximate solutions exist. See \cite{Cohen06} 
as an example.
\item Fairness requirement: Here the problem is similar to the partitioning problem 
described in Section \ref{subsec:clustering}, with the additional requirement that 
the resulted clusters are balanced, in the sense that the interference experienced 
by cellular users due to D2D users are \textit{almost} equal. Formally, desired is 
to solve (\ref{eq:Second}), subject to (\ref{eq:clustConOne}), (\ref{eq:clustConTwo}) 
and (\ref{eq:clustConThree}), so that $\sum_{l \in \mathcal{L}_{1}}\sum_{k \in \mathcal{K}_{1}}
w_{kl}\approx \sum_{l \in \mathcal{L}_{2}}\sum_{k \in \mathcal{K}_{2}}w_{kl}\approx...
\approx \sum_{l \in \mathcal{L}_{Q}}\sum_{k \in \mathcal{K}_{Q}}w_{kl}$. It should 
be emphasized that in this context, the burden of D2D communication is divided (almost) 
equally among cellular users, which does not necessarily result in achieving equal 
utilities by all of them.
\end{itemize}
\section{Power Control}\label{sec:Power}
This section deals with the second step of resource assignment, i.e., D2D power control, 
which aims at optimizing the performance of D2D users. 
\subsection{Power Control Game}\label{subsec:PCGameDef}
As described in the foregoing section, while performing the channel assignment, 
the BS ignores the potential interferences that might arise among D2D users, due 
to the lack of information and also their lower priority. In essence, D2D users 
are partitioned into clusters and each cluster is assigned a single channel. Given 
no information, each D2D user therefore intends to maximize its own utility, thereby 
causing interference to the users with whom it shares a channel. By power control, 
however, interference can be managed so that the channel assigned to each cluster 
is utilized efficiently. We model the power control problem as a game with incomplete 
information, defined on a discrete strategy set. We show that the game is potential 
and characterize the set of Nash equilibria. To this end, we define (exact) potential 
games \cite{Monderer96} and Nash equilibrium \cite{Nash51}.
\begin{definition}[Potential Game]
\label{de:PG}
Consider a strategic game $\mathfrak{G}=\left\{\mathcal{K},
\mathcal{I},\left\{R_{k}\right\}_{k \in \mathcal{K}}\right\}$, where $\mathcal{K}$ 
is the set of $K$ players, $\mathcal{I}$ is the set of pure-strategy joint action 
profiles of all players, and $R_{k}:\mathcal{I} \rightarrow \mathbb{R}^{+}$ denotes 
the payoff function of player $k$. Then $\mathfrak{G}$ is an exact potential game 
if there exists a function $v: \mathcal{I} \rightarrow \mathbb{R}^{+}$ such that 
for all $k \in \mathcal{K}$ we have
\begin{equation}
\label{eq:PG}
R_{k}(i_{k},\mathbf{i}_{-k})-R_{k}(i'_{k},\mathbf{i}_{-k})=v(i_{k},\mathbf{i}_{-k})
-v(i'_{k},\mathbf{i}_{-k}),
\end{equation}
where $i_{k}$ is the action of player $k$ while $\mathbf{i}_{-k}$ denotes the joint 
action profile of all players except for $k$. Any such function $v$ is called 
a potential of $\mathfrak{G}$.
\end{definition}
\begin{definition}[Nash equilibrium]
\label{de:Nash}
A joint strategy profile $\mathbf{i}=(i_{1},...,i_{k},...,i_{K})$ is called a pure-strategy Nash 
equilibrium if for all $k \in \mathcal{K}$ and all actions $i'_{k}$, the joint strategy 
profile $\mathbf{i}'= (i_{1},...,i'_{k},...,i_{K})$ yields $R_{k}(\mathbf{i}')\leq 
R_{k}(\mathbf{i})$.
\end{definition}
As clusters are assigned orthogonal channels, the actions of D2D users inside 
any given cluster do not affect the utilities of the users outside that cluster. 
Therefore the power allocation problem in any cluster $q \in \left\{1,...,Q\right\}$ 
can be defined as a game among $K_{q}$ D2D users.
\begin{definition}[Cluster Power Allocation Game]
\label{de:CPAG}
The power allocation game of cluster $q \in \left\{1,...,Q\right\}$ is a 
strategic game defined as $\mathfrak{G_{q}}=\left\{\mathcal{K}_{q},
\mathcal{I},\left\{R_{k} \right\}_{k \in \mathcal{K}_{q}}\right\}$, where 
$\mathcal{K}_{q}$ is the set of D2D users assigned to channel $q$, $\mathcal{I}=
\bigotimes_{k=1}^{K_{q}}\left \{p_{d}^{(1)},p_{d}^{(2)},...,p_{d}^{(M)}\right\}$ 
is the set of joint actions with realizations $\mathbf{p}_{d,q}=(p_{1},...,p_{K_{q}})$, 
and $R_{k}:\mathcal{I} \rightarrow \mathbb{R}^{+}$ is the payoff function of 
player $k \in \left\{1,...,K_{q}\right\}$ defined in (\ref{eq:DDThroughput}) 
(Section \ref{subsubsec:Transmission}). 
\end{definition}
The main difference between the cluster power allocation game and the 
standard power control games investigated in other studies including 
\cite{Scutari06} is that the strategy set of players is here extracted 
from a \textit{discrete space}, while in the previous contributions the 
strategy space is continuous. Consequently, most of the existing results 
do not hold, and hence we proceed to the following theorem.
\begin{theorem}
\label{th:PG}
\begin{enumerate}[a)]
\item The cluster power allocation game (Definition \ref{de:CPAG}) is an 
exact potential game with potential
\begin{equation}
\label{eq:potentialFunc}
v(\mathbf{p}_{d,q})=\sum_{k \in \mathcal{K}_{q}}\log\left(p_{k}\right)-
\sum_{k \in \mathcal{K}_{q}}c p_{k}. 
\end{equation} 
\item Denote the set of potential maximizers by $\mathcal{V}_{\textup{max}}$. Then, 
a joint action profile $\mathbf{p}_{d,q}$ is a Nash equilibrium if and only if 
$\mathbf{p}_{d,q} \in \mathcal{V}_{\textup{max}}$. 
\end{enumerate}  
\end{theorem}
\begin{IEEEproof}
See Appendix \ref{sec:AppTwo}.
\end{IEEEproof}
\subsubsection{Quality of Service Guarantee}\label{subsec:QoSDD}
In Definition \ref{de:CPAG}, we assume that D2D users have no 
strict QoS requirement, and only aim at maximizing some reward, expressed in 
terms of SIR and cost. As a result, the set of joint strategies yields $\mathcal{I}
=\bigotimes_{k=1}^{K_{q}}\left \{p_{d}^{(1)},p_{d}^{(2)},...,p_{d}^{(M)}\right\}$. 
While this formulation holds for many problems, there are some cases where D2D 
users need to meet some specific QoS requirements, expressed for instance in terms 
of some minimum SIR value. In such scenarios, each player tries to selfishly solve 
the following problem
\begin{equation}
\label{eq:opti}
\underset{p_{k}\in\mathcal{A}_{k} \left(\mathbf{p}_{-k} \right )}{\textup{minimize}~~p_{k}}
\end{equation}
where $\mathcal{A}_{k}$ is the set of strategies for player $k$, which depends
on the joint strategy profile of its opponents, $\mathbf{p}_{-k}$, and is given 
by
\begin{equation}
\label{eq:strSet}
\mathcal{A}_{k}=\left \{ p_{k} \in \mathcal{M}:\gamma_{k}\geq \Gamma_{k}\right \}.
\end{equation}
Here $\Gamma_{k}$ is the minimum required SIR for D2D user $k$ to meet its QoS 
target. In other words, the players' strategy sets are correlated so that any 
player plays only the actions that satisfy its QoS constraint, given the actions 
of opponents. It is known that the problem in (\ref{eq:opti}) can be modeled as 
a strategic game, where the utility of each player $k$ is defined as $R_{k}=-p_{k}$ 
\cite{Chiang07} or $R_{k}=-\log \left(p_{k}\right)$ \cite{Scutari06}. Along similar 
lines with Theorem \ref{th:PG}, it is straightforward to show that the game is an 
exact potential game with potential $v(\mathbf{p}_{d,q})=\sum_{k=1}^{K_{q}}
R_{k}(\mathbf{p}_{d,q})$, provided that the original problem (\ref{eq:opti}) is 
feasible. 
\subsection{Q-Learning Better-Reply Dynamics}\label{subsec:FPQGame}
According to the system model, in the cluster power allocation game (Definition 
\ref{de:CPAG}), the utility functions are not known by players (D2D users) in 
advance. Therefore they require interacting with the environment in order to 
i) learn the reward functions, and ii) achieve equilibrium. We consider the cluster 
power allocation game to be a game with noisy payoffs. In such games, for each 
joint action profile $\textbf{i}\in \mathcal{I}$ of $K_{q}$ players, the utility 
achieved by player $k$ at each interaction can be written as $R_{k}=\bar{R}_{k}
(\textbf{i})+ e_{k}$, where $\bar{R}_{k}$ is the true expected value of the utility 
function $R_{k}$ and $e_{k}$ is a random fluctuation with zero mean and bounded 
variance, independent from all other random variables. During the learning process, 
each player faces a trade-off between gathering information (learning) on the one 
hand and using information to achieve higher utility (control) on the other hand. 
This trade-off is known as exploration-exploitation dilemma. In order to deal 
with this dilemma and also to achieve equilibrium in a distributed 
manner, we use Q-learning better-reply dynamics \cite{Chapman13}. This strategy 
consists of three main steps that are performed recursively: 1) Observe 
the personal reward and also the actions of opponents.\footnote{When using multi-agent 
Q-learning algorithms, conventionally it is assumed that every agent observes the 
state of the environment and/or the actions of its opponents \cite{Vlachogiannis04}. 
In our model, players are therefore required to announce their transmit powers, 
for example by broadcasting in a specific time period, borrowed from the total 
transmission time. This overhead, however, is much less than that of the frequent 
and pairwise data exchange, for which usually a control channel is allocated 
\cite{Liu12}. The reason is that after convergence, which is achieved relatively 
fast, the transmit powers of players remain fixed. Therefore no more broadcasting 
is required and the borrowed time period is again available for data transmission. 
We also assume that the players have a finite memory of length $m$; that is, at each 
trial, each player remembers the played joint action profiles of exactly $m$ past trials.}~2) 
Update the Q-values of the played joint action profile. 3) With a small probability, 
$\epsilon \ll 1$, select an action uniformly at random, while with a large probability, 
$1-\epsilon$, play according to the better-reply dynamics that is described in the 
following definition.
\begin{definition}[Better-Reply Dynamics \cite{Chapman13}]
\label{def:BR}
Assume that at some trial $t-1$, a player $k$ plays with action $p_{k,t-1}$. Then, at trial $t$, 
with probability $\zeta_{k}$, the player selects the same action as in the previous trial, $t-1$, 
i.e., $p_{k,t}=p_{k,t-1}$. With probability $1-\zeta_{k}$, however, the player selects an action 
according to a distribution that puts positive probabilities only on actions that are better 
replies to its (finite) memory than $p_{k,t-1}$. For instance, it selects an action according a 
uniform distribution over all better-replies.
\end{definition}
For readers' convenience, the detailed strategy is described in Algorithm \ref{alg:FPQGame} 
for some player $k \in \mathcal{K}_{q}$.
\begin{algorithm}
\caption{Q-Learning Better-Reply Dynamics \cite{Chapman13}}
\label{alg:FPQGame}
\small
\begin{algorithmic}[1]
\STATE Select arbitrary positive constants $c_{\lambda}$ and $c_{\varepsilon}$.
\STATE Select learning parameters $\rho_{\lambda}\in \left[\frac{1}{2},1\right]$.
\STATE Let $\delta_{k,t}$ be the mixed strategy of player $k$ at time $t$. Let 
       $\delta_{k,1}$ be the uniform distribution over all actions (power levels). 
\STATE Select an action, $p_{k,t}$, using $\delta_{k,1}$. Play and observe the reward.
\FOR {$t=2,...,T$} 
     \STATE Let
       \begin{equation}
            \label{eq:varepsilon}
            \varepsilon_{t}=c_{\varepsilon}t^{\frac{-1}{K_{q}}}.   
       \end{equation}
     \STATE 
       \begin{itemize}    
         \item With probability $\varepsilon_{t}$, let $\delta_{k,t}$ be the uniform 
               distribution over all actions.
         \item With probability $1-\varepsilon_{t}$, perform the following (better-reply dynamics):
            \begin{itemize}
              \item With probability $\zeta_{k}$, let $\delta_{k,t}$ be the Dirac 
                    probability distribution on $p_{k,t-1}$.
              \item With probability $1-\zeta_{k}$, let $\delta_{k,t}$ be the uniform 
                    distribution over all actions that are better replies to the 
                    full (finite) memory than $p_{k,t-1}$.
            \end{itemize}
      \end{itemize}       
    \STATE Using $\delta_{k,t}$, select the action of time $t$, $p_{k,t}$, and play.     
    \STATE Announce the selected action. Moreover, observe the played joint action profile of 
           other players, $\mathbf{p}_{-k,t}$, and also the achieved reward, $R_{k}(\mathbf{p}_{d,q}^{(t)})$, 
           where $\mathbf{p}_{d,q}^{(t)}=(p_{k,t},\mathbf{p}_{-k,t})=\left(p_{1,t},..,p_{k,t},..,p_{K_{q},t} \right)$ .  
    \STATE Update the Q-value of the played joint action profile as
           \begin{equation}
           Q_{k,t+1}(\mathbf{p}_{d,q}^{(t)})=Q_{k,t}(\mathbf{p}_{d,q}^{(t)})+\lambda_{t}\left (R_{k}                                                                          (\mathbf{p}_{d,q}^{(t)})-Q_{k,t}(\mathbf{p}_{d,q}^{(t)})\right)\mathbf{1}_{\mathbf{p}_{d,q}^{(t)}},
           \end{equation}
           with 
           \begin{equation}
           \label{eq:lambda}
           \lambda_{t}=\left(c_{\lambda}+\#^{t}[\mathbf{p}_{d,q}^{(t)}]\right)^{-\rho_{\lambda}},
           \end{equation}
           where $\#^{t}[\mathbf{p}_{d,q}^{(t)}]$ denotes the number of trials in which $\mathbf{p}_{d,q}^{(t)}$ 
           is played while $\mathbf{1}_{\mathbf{p}_{d,q}^{(t)}}$ is the indicator function.    
\ENDFOR
\end{algorithmic}
\end{algorithm}
\begin{theorem}[\cite{Chapman13}]
\label{th:Qlearning}
The Q-learning better-reply dynamics (Algorithm \ref{alg:FPQGame}), with 
$\varepsilon^{t}$ and $\lambda^{t}$ given by (\ref{eq:varepsilon}) and 
(\ref{eq:lambda}) respectively, converges to a pure Nash equilibrium in 
games with noisy unknown rewards that are generic and admit a potential 
function.
\end{theorem}
\begin{corollary}
By using Q-learning better-reply dynamics, the cluster power allocation game 
(Definition \ref{de:CPAG}) converges to a pure Nash equilibrium that maximizes 
the potential function.
\end{corollary}
\begin{IEEEproof}
The proof directly follows from Theorem \ref{th:PG} and Theorem \ref{th:Qlearning}.
\end{IEEEproof}
\begin{remark}
Let $\alpha=O\left(M^{K_{q}}\right)$ be the size of the normal form 
representation of the cluster power allocation game. Similar to any other equilibrium-learning strategy, 
Algorithm \ref{alg:FPQGame} follows a better-reply path to a pure Nash 
equilibrium, whose length grows exponentially in $\alpha$ \cite{Christo'13}. 
On the other hand, as for Q-learning, the Q-value of all joint action profiles 
(that is equal to $\alpha$) must be learned. As a result, the running time is 
at least exponential in the size of the game, i.e., $O(c^{\alpha})$ for some 
constant $c>1$. Thus, for a specific number of players (which is determined by 
clustering), smaller $M$ (number of power levels) yields faster convergence, as 
one expects intuitively. Similarly, smaller $M$ yields lower computational complexity.
\end{remark}
\begin{remark}
As described before, in any game, complexity and convergence speed 
to equilibrium depends dramatically on the size of the game. This dependency becomes 
even stronger for games with incomplete information, as the reward of all joint action 
profiles must be learned through successive interactions. As a result, it is of utmost 
importance to reduce the size of the game and/or to use any available information. The 
designed two-stage resource allocation mechanism strictly follows this policy, as by 
excluding cellular users from the set of players, and channels from the set of actions, 
the game size reduces abruptly in comparison with a one-stage game, while the available 
information at the BS is used efficiently. Additionally, it allows taking the priority 
of cellular users into account, which is not possible in a one-stage game.
\end{remark}
\subsection{Efficiency of Equilibrium}\label{subsec:EofE}
According to Theorem \ref{th:Qlearning}, for the cluster power 
allocation game, any pure-strategy Nash equilibrium maximizes the potential 
function, given by (\ref{eq:potentialFunc}). It should be however noted that here 
the potential function is not equal to social welfare, $f(\mathbf{p}_{d,q})
=\sum_{k=1}^{K_{q}}R_{k}(\mathbf{p}_{d,q})$. Therefore, the pure-strategy Nash 
equilibrium does not necessarily maximizes the sum utilities of all players, 
although such a solution is desired. The inefficiency of equilibrium is formalized 
by \textit{price of stability}, defined below. 
\begin{definition}[Price of Stability \cite{Anshelevich04}]
\label{def:PoS}
Let $f(\mathbf{p}_{d,q})$ be an objective function such as social welfare, which 
we wish to maximize. Moreover, let $\mathcal{N}$ denote the set of pure 
Nash equilibriums of the cluster power allocation game. Then the price of stability 
(PoS) is defined as
%
%
\begin{equation} 
\label{eq:PoS}
\textup{PoS}=\frac{\max f(\mathbf{p}_{d,q})}{\underset{\mathbf{p}_{d,q} \in \mathcal{N}}
{\max}f(\mathbf{p}_{d,q})}.
\end{equation}
\end{definition}
Note that the objective function to be optimized and the solution set being evaluated 
might vary. For instance, the objective function could be the minimum reward (so that 
the optimization problem corresponds to max-min fairness criterion), or the set of solution might 
also include mixed-strategy equilibria. The following proposition provides an upper-bound 
for the inefficiency of pure-strategy Nash equilibrium in the cluster power allocation 
game.
\begin{proposition}
\label{pr:PoSBound}
For the cluster power allocation game described in Definition \ref{de:CPAG}, define  
\begin{equation} 
\label{eq:GamMin}
\gamma_{\min}:=\underset{k \in \mathcal{K}}{\min}~\frac{p_{d}^{(1)}h_{kk',q}}{1+\sum_{j 
\in \mathcal{K}_{q},j\neq k}p_{d}^{(M)}h_{jk',q}+p_{c}h_{bk',q}}.
\end{equation}
Then we have $1\leq \textup{PoS}\leq \frac{\log \left (p_{d}^{(M)}\right)}
{\log\left(\gamma_{\textup{min}}\right)}$.
\end{proposition}
\begin{IEEEproof}
See Appendix \ref{sec:AppFive}.
\end{IEEEproof}
Although the bound provided by Proposition \ref{pr:PoSBound} is loose, in general 
it clearly shows that a larger $p_{d}^{(M)}-p_{d}^{(1)}$ value of (range of the set 
of power levels, $\mathcal{M}$) may yield higher inefficiency of pure Nash equilibrium. 
Recall that large range of $\mathcal{M}$ has also an adverse effect on the lower-bound 
given by (\ref{eq:lowerBound}). Therefore the two-stage resource allocation mechanism 
is particularly suitable for $\mathcal{M}$ with small ranges. It is worth mentioning 
that for games with multiple equilibriums, the inefficiency of the worst Nash equilibrium 
is formalized by \textit{price of anarchy} (PoA) \cite{Koutsoupias13}. Calculating PoA 
is mathematically involved and lies out of the scope of this paper.
\section{Numerical Analysis}\label{sec:Numerical}
\begin{figure}[t]
\centering
\includegraphics[width=0.62\textwidth]{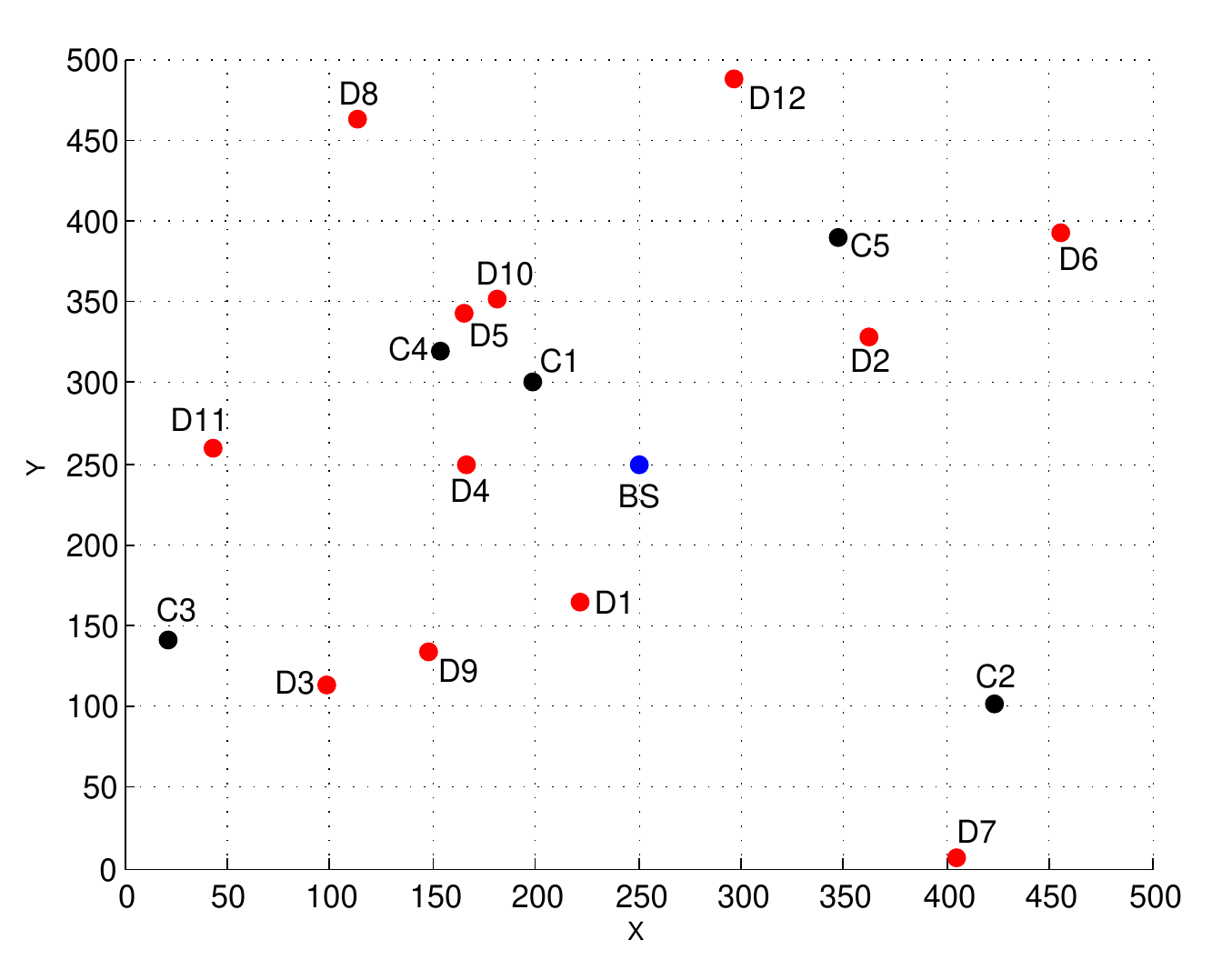}
\caption{Network model consisting of D2D transmitters (D$i$, 
$i \in \left \{1,...,12\right \}$) and cellular receivers (C$i$, 
$i \in \left \{1,...,5\right \}$).}
\label{Fig:SysModel}
\end{figure}

We consider an underlay D2D communication system, consisting of twelve D2D users 
($K=12$) and five cellular users ($L=5$), as depicted in Figure \ref{Fig:SysModel}. 
Note that only the \textit{transmitter side} of D2D users are shown in the figure, 
as receivers do not cause any interference to cellular users and therefore do not 
impact the channel allocation (see also the definition of estimated network graph 
in Section \ref{subsec:clustering}). Also note that for numerical analysis, the 
locations of cellular and D2D users, as well as channel gains, are selected randomly. 
According to the system model (Section \ref{subsubsec:Network}), there exist five 
orthogonal channels ($Q=5$). Each D2D user $k \in \mathcal{K}$ selects a transmit 
power from the set of power levels, $\mathcal{M}=\left \{2,4 \right \}$. Moreover, 
the transmit power of the BS to the cellular users is $p_{c}=7$.
\subsection{Channel Allocation}\label{subsec:ChannelAllocation}
Table \ref{Tb:BMGain} includes $h_{bl,q}$ (cellular-BS average channel gains) 
for $l,q \in \left \{1,...,5\right\}$, which is assumed to be known by the BS 
together with the network topology (Figure \ref{Fig:SysModel}), according to 
Assumption \ref{as:AvailInfo} (Section \ref{sec:SystemModel}). Based on this 
information and by using the graph-theoretical channel allocation scheme described 
in Section \ref{sec:Channel}, the BS assigns each (cellular and D2D) user a channel, 
as summarized in Table \ref{Tb:Channel}. Based on Table \ref{Tb:BMGain} and Figure 
\ref{Fig:SysModel}, it can be concluded that by the channel allocation given in 
Table \ref{Tb:Channel}, both (\ref{eq:First}) and (\ref{eq:Second}) are satisfied. 
\begin{table}[ht]
\caption{BS to cellular average channel gains}
\label{Tb:BMGain}
\begin{center}
  \begin{tabular}{|c|c|c|c|c|c|}
  \hline
   \backslashbox{User}{Channel} &  1   &  2   &  3   &  4   &  5    \\ \hline 
   C$1$                         & 0.04 & 0.01 & 0.27 & 0.12 & 0.04  \\ \hline
   C$2$                         & 0.29 & 0.06 & 0.15 & 0.18 & 0.26  \\ \hline
   C$3$                         & 0.31 & 0.46 & 0.24 & 0.19 & 0.06  \\ \hline
   C$4$                         & 0.12 & 0.06 & 0.29 & 0.34 & 0.16  \\ \hline
   C$5$                         & 0.24 & 0.08 & 0.23 & 0.41 & 0.07  \\ \hline
  \end{tabular}
 \end{center} 
\end{table}  

As discussed in Section \ref{subsec:Remarks}, it is also possible to change the 
criterion of channel allocation from maximizing the social welfare to address the 
QoS guarantee or fairness issues (of cellular users). Assume that the required QoS 
of any cellular user $l \in \mathcal{L}$ is satisfied if it achieves some minimum 
utility, say $R_{l,\textup{min}}=3.5$.\footnote{Note that the QoS requirements of 
cellular users do not need to be necessarily similar.}~Therefore by using the data 
given in Table \ref{Tb:BMGain}, the maximum tolerable interference of each cellular 
user can be simply calculated. A channel allocation that guarantees the QoS satisfaction 
of all cellular users is summarized in Table \ref{Tb:ChannelQ}. Moreover, the result 
of channel assignment based on fairness among cellular users is shown in Table \ref{Tb:ChannelF}.\footnote{Note 
that the solutions are approximately-optimal and also not unique.}
\begin{table}[ht]
\label{}
\centering
\caption{Channel allocation based on different performance criteria of cellular users}
\subtable[Maximum Aggregate Utility]{
  \begin{tabular}{ |c|c|}
  \hline
  Channel &     User                          \\ \hline 
   $1$    &  C$5$,D$1$,D$3$,D$9$              \\ \hline
   $2$    &  C$3$,D$2$,D$6$,D$7$,D$12$        \\ \hline
   $3$    &  C$1$                             \\ \hline
   $4$    &  C$4$                             \\ \hline
   $5$    &  C$2$,D$4$,D$5$,D$8$,D$10$,D$11$  \\ \hline
  \end{tabular}
  \label{Tb:Channel}
}
\quad
\subtable[QoS guarantee]{
  \begin{tabular}{ |c|c|}
  \hline
  Channel &     User                          \\ \hline 
   $1$    &  C$5$,D$3$,D$9$                   \\ \hline
   $2$    &  C$3$,D$1$,D$2$,D$11$,D$12$       \\ \hline
   $3$    &  C$1$,D$8$,D$10$                  \\ \hline
   $4$    &  C$4$,D$6$                        \\ \hline
   $5$    &  C$2$,D$4$                        \\ \hline
  \end{tabular}
  \label{Tb:ChannelQ}
}  
\quad
\subtable[Fairness]{
  \begin{tabular}{ |c|c|}
  \hline
  Channel &     User                          \\ \hline 
   $1$    &  C$5$,D$3$,D$9$,D$11$             \\ \hline
   $2$    &  C$3$,D$2$,D$12$                  \\ \hline
   $3$    &  C$1$,D$1$,D$8$                   \\ \hline
   $4$    &  C$4$,D$6$,D$7$                   \\ \hline
   $5$    &  C$2$,D$4$,D$5$,D$10$             \\ \hline
  \end{tabular}
  \label{Tb:ChannelF}
}
\end{table}

\begin{figure}[t]
\centering
\includegraphics[width=0.55\textwidth]{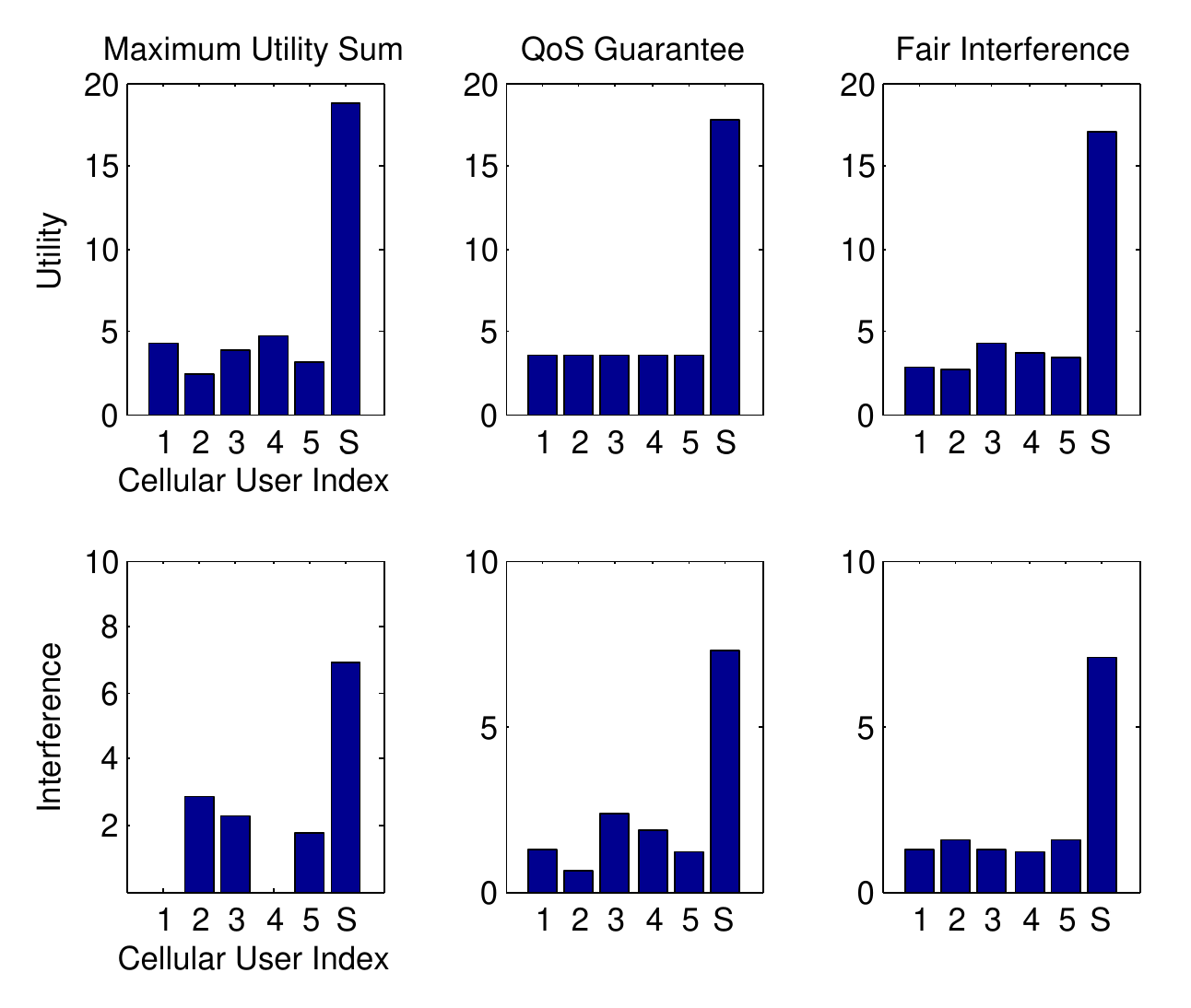}
\caption{Average utility and interference experienced by cellular users under three 
criteria (S:Sum).}
\label{Fig:RewCel}
\end{figure}

The achieved average rewards of cellular users under all three criteria are shown 
in Figure \ref{Fig:RewCel}. It can be seen that to achieve the highest utility sum, 
some cellular users do not experience any interference, while some others are strongly 
disturbed. In case of QoS guarantee, users with higher channel gains experience more 
interference and vice versa, so that at the end all cellular users are satisfied. 
Moreover, by Table \ref{Tb:ChannelQ}, in the current setting, all D2D users can be 
served without violating the QoS requirement of cellular users.\footnote{Clearly, 
this might not be always the case. In fact, given a specific QoS requirement of 
cellular users, the number of D2D users that can be served  depends strongly on 
network topology, channel quality and the required QoS.}~In the last criterion, 
all cellular users experience almost equal amounts of interference, regardless of 
their achieved utilities. 

\begin{figure}[t]
\centering
\includegraphics[width=0.44\textwidth]{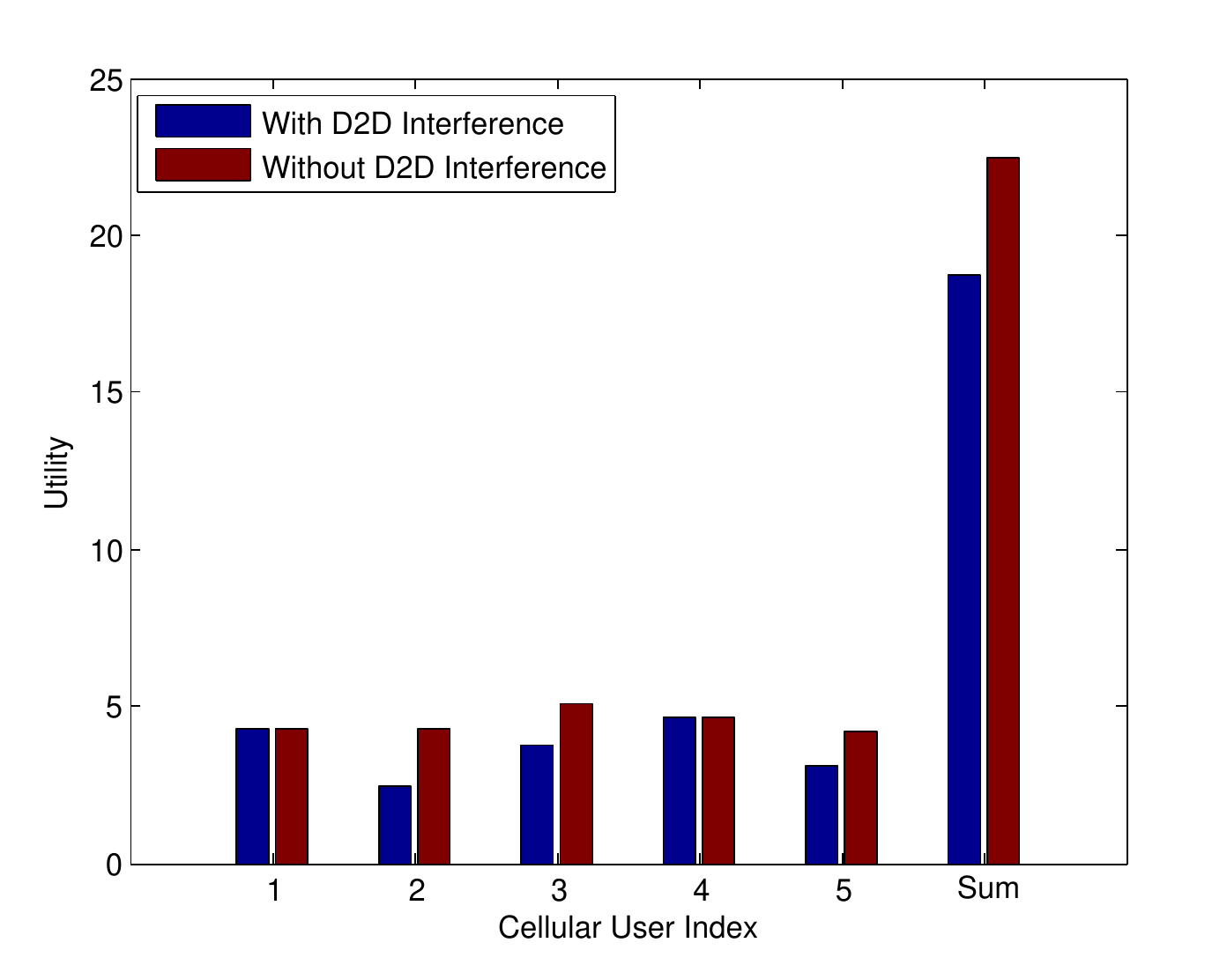}
\caption{Performance loss of cellular users due to channel sharing with the 
allocation criterion being the maximization of cellular utility sum.}
\label{Fig:PerLos}
\end{figure}

For our primary channel allocation criterion, i.e., maximizing the aggregate utility of 
cellular users, it is of interest to investigate the performance loss of cellular users, 
caused by sharing resources with D2D users. The performance degradation is shown in 
Figure \ref{Fig:PerLos}, where the achievable utilities of cellular users without any 
interference (no channel sharing) are shown in comparison with the case where \textit{all} 
D2D users are assigned some channel. From this figure, it can be concluded that in the 
current setting, serving all D2D users costs approximately 15\% performance loss to cellular 
users.
\subsection{Power Control}\label{subsec:PowerControl}
\begin{figure}[t]
\centering
\includegraphics[width=0.38\textwidth]{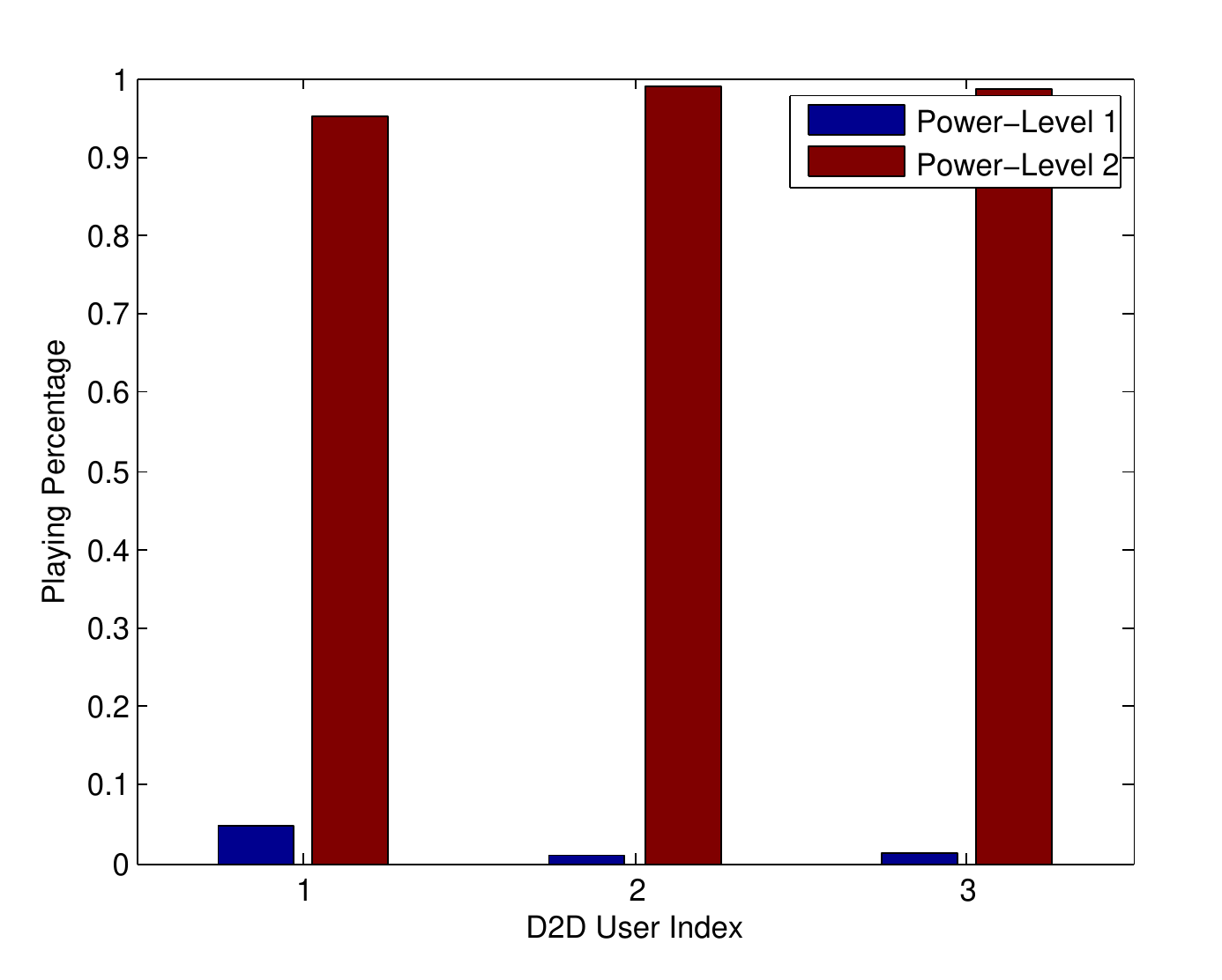}
\caption{Fraction of trials in which any given action is played by D2D users.}
\label{Fig:Act}
\end{figure}

\begin{figure}[t]
\centering
\includegraphics[width=0.58\textwidth]{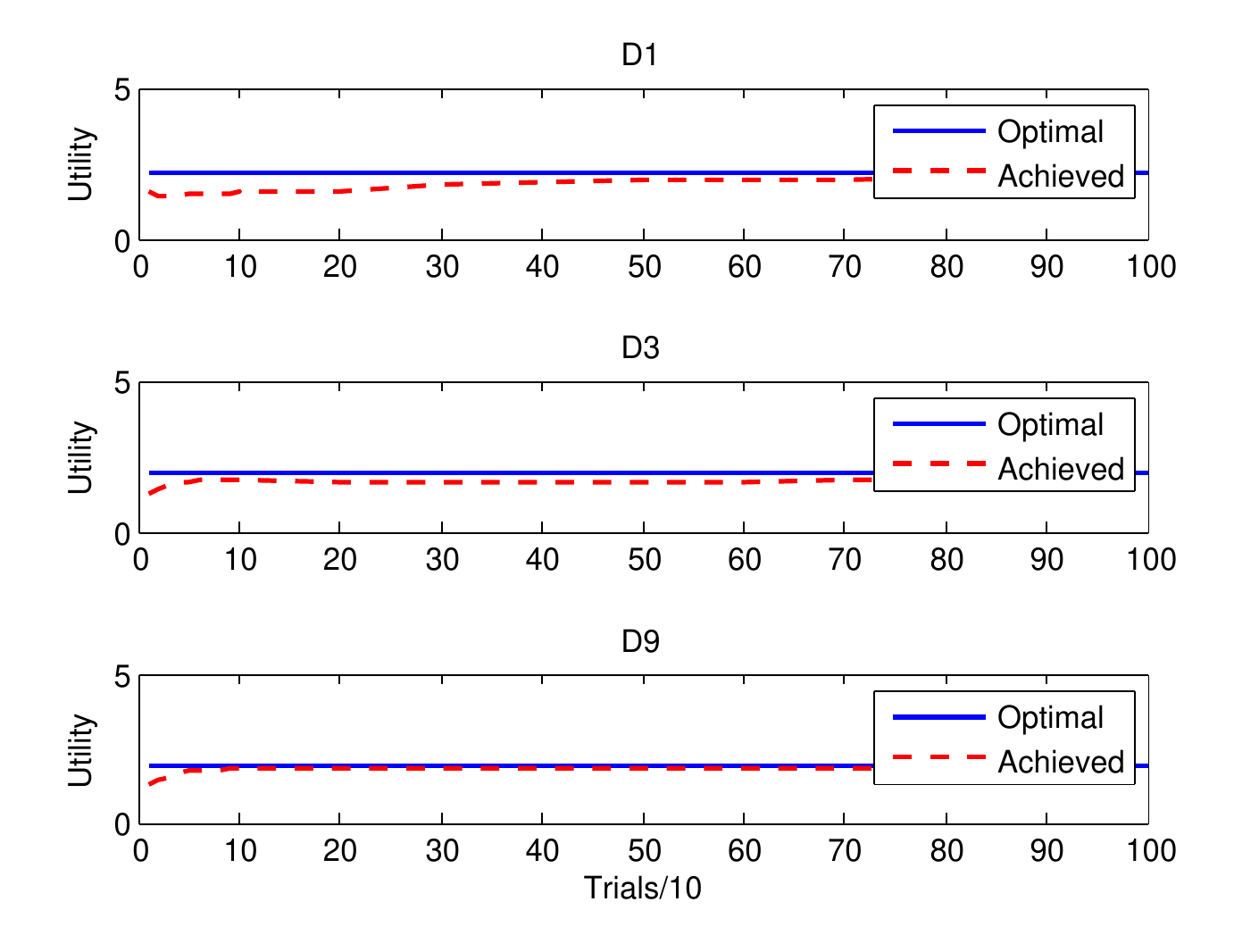}
\caption{Utilities achieved by D2D users versus utility values at equilibrium.}
\label{Fig:RewDD}
\end{figure}

From Table \ref{Tb:Channel}, it can be observed that minimum-weighted 
partitioning divides the D2D and cellular users into five clusters, each 
allocated a frequency channel. In this section, we investigate the power 
control game of the first cluster, i.e., the cluster that includes three 
D2D users (D$1$, D$3$ and D$9$) and is assigned channel one. The games of 
other clusters are similar. The game horizon and price factor are considered 
to be $T=2\times 10^{3}$ and $c=0.1$, respectively. The joint action profiles 
of the three users as well as their average rewards are given by Table \ref{Tb:JointRew}. 
\begin{table}[ht]
\caption{Joint Reward Table}
\label{Tb:JointRew}
\begin{center}
  \begin{tabular}{ |c|c||c|c| }
  \hline
    Joint Action  & Joint Reward         & Joint Action  & Joint Reward       \\ \hline 
    $(2,2,2)$     & $(2.60,2.36,2.10)$   & $(4,4,2)$     & $(2.80,2.54,0.30)$ \\ \hline
    $(2,4,2)$     & $(1.80,3.36,1.30)$   & $(2,4,4)$     & $(1.22,2.54,2.28)$ \\ \hline
    $(4,2,2)$     & $(3.58,1.56,1.28)$   & $(4,2,4)$     & $(2.80,0.98,2.28)$ \\ \hline
    $(2,2,4)$     & $(1.80,1.56,3.08)$   & $(4,4,4)$     & $(2.20,1.98,1.90)$ \\ \hline  
    \end{tabular}
  \end{center} 
\end{table}  
From this table, the action profile $(4,4,4)$, i.e., $(p_{d}^{(2)},
p_{d}^{(2)},p_{d}^{(2)})$, is the unique Nash equilibrium, which maximizes 
the potential function. Hence the game converges theoretically to this 
point. Figure \ref{Fig:Act} describes the frequency in which any given 
action is played by each D2D user. It can be seen that the equilibrium 
strategy is played almost all the time. 
Figure \ref{Fig:RewDD} depicts the average utility of D2D users versus the 
equilibrium reward, confirming that in a short time the average reward of 
every player converges to that of equilibrium point. 
\subsection{Overall Performance}\label{subsec:Overall}
\begin{figure}[t]
\centering
\includegraphics[width=0.53\textwidth]{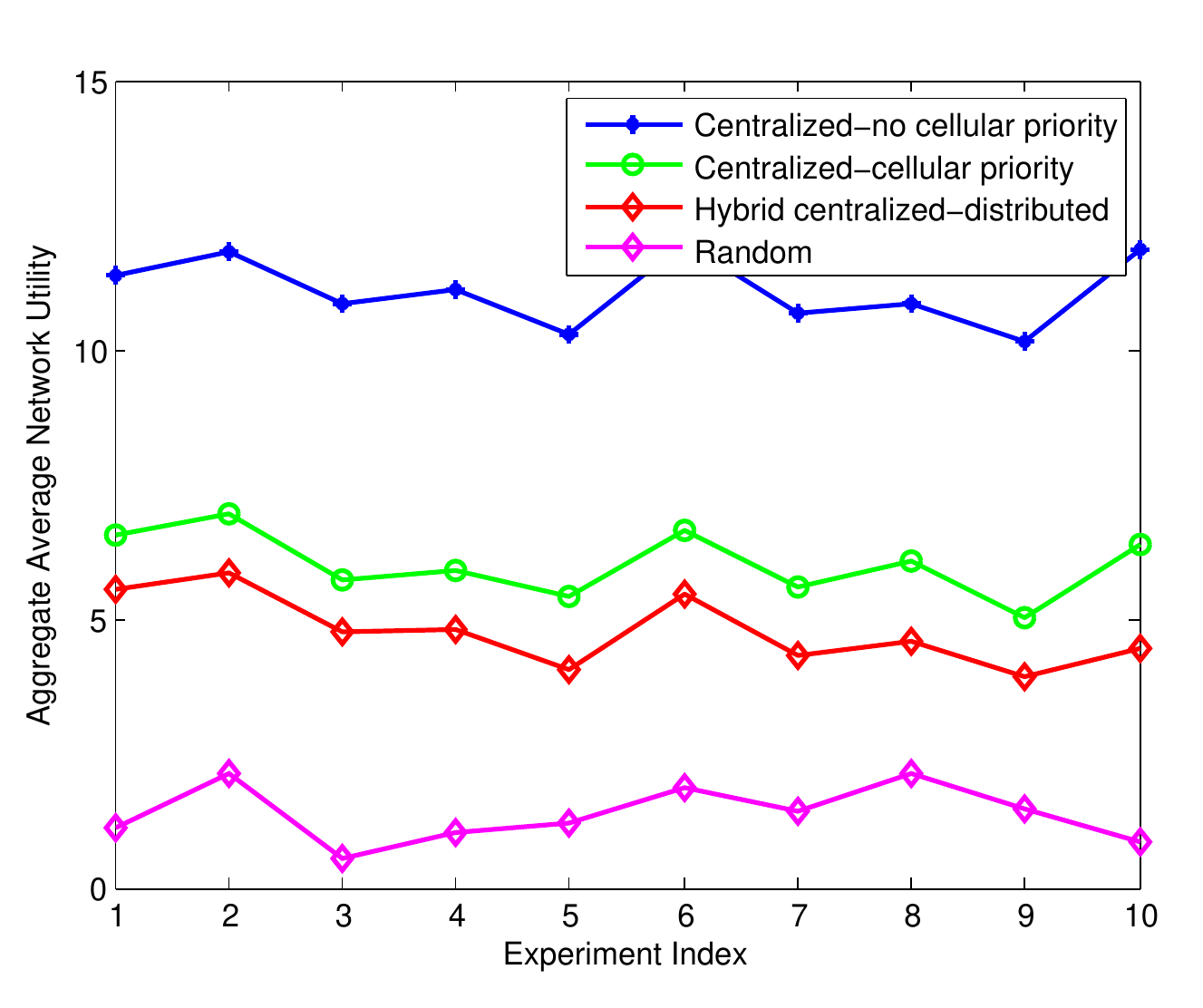}
\caption{Overall performance of the proposed scheme compared to some other strategies.}
\label{Fig:Comp}
\end{figure}

In order to evaluate the overall performance of the proposed resource allocation 
scheme, we compare it with three other strategies that are described below.
\begin{itemize}
\item Centralized approach that is based on the exhaustive search given global information. In 
      accordance with the concept of underlay D2D networks, the priority is here granted to the 
      cellular users. Formally, the selected joint channel and power allocation vector maximizes 
      $\sum_{l=1}^{L}R_{l}$, and ties are broken in favor of the allocation vector that yields 
      higher aggregate D2D utility, i.e., larger $\sum_{k=1}^{K}R_{k}$. 
\item Centralized approach that is based on the exhaustive search given global information, but 
      \textit{without} considering the priority of cellular users. Formally, the algorithm 
      searches for the joint channel and power allocation vector that maximizes $\sum_{l=1}^{L}R_{l}
      +\sum_{k=1}^{K}R_{k}$.       
\item Random resource allocation, where the channel and power levels are assigned 
      using uniform distribution.
\end{itemize}
As applying the exhaustive search approach to the large network investigated before 
(Figure \ref{Fig:SysModel}) yields excessive complexity ($5^{16} \times 2^{12}$ cases 
should be searched), we turn to a smaller network with $L=Q=M=2$ and $K=6$. Ten experiments 
are performed. For \textit{each experiment}, independent from others, average channel 
gains and users' locations are selected randomly. In other words, ten random simulation 
settings are selected. For \textit{each experiment}, the sum of average rewards of all 
(cellular and D2D) users is simulated over $T=10^{3}$ trials. Results are depicted in 
Figure \ref{Fig:Comp}.
From this figure, it can be concluded that the utility achieved by our proposed resource 
allocation scheme is almost equal to the highest possible aggregate network utility when 
taking the priority of cellular users into account. Note that the difference is due to i) 
bounding and decomposition techniques that are used in Section \ref{sec:Channel}, and ii) 
the inefficiency of equilibrium that is described in Section \ref{sec:Power}. Hence the 
performance gap is in fact the cost of i) absence of a coordinator, ii) lack of information, 
and iii) low time and computational complexity tolerance. It is also worth noting that larger 
network utility sum can be achieved by neglecting cellular priority; nevertheless, such 
setting does not comply with the concept of underlay D2D communication, since cellular 
users might be extremely disturbed. It is also worth mentioning that for larger number 
of D2D and cellular users, the number of possible channel and power allocation vectors 
grows exponentially, and hence centralized resource allocation based on exhaustive search 
yields excessive cost in terms of time and computational complexity, as well as a large 
overhead that is required for information acquisition. Our approach, in contrast, offers 
low complexity and overhead; hence it is specifically suitable for large networks.

\section{Conclusion and remarks}\label{sec:conclusion}
We studied an underlay D2D communication system, and proposed a two-stage 
resource allocation strategy that takes the priority of cellular users into 
account, and relies on strictly limited information. In the first stage, 
centralized channel allocation is performed by using a graph-theoretical 
method. The method offers high flexibility for selecting the allocation 
criteria, for instance aggregate utility, fairness or QoS guarantee. The 
complexity was shown to be polynomial in the number of users. In the second 
stage, power control problem is modeled as a game with incomplete information. 
We showed that the game is an exact potential game defined on a discrete 
strategy set, and therefore Q-learning better-reply dynamics can be used 
by players to achieve a pure strategy Nash equilibrium in a distributed manner. 
The set of Nash equilibria was shown to be equivalent to the set of potential 
maximizers, and the inefficiency of Nash equilibrium was discussed. Extensive 
numerical analysis demonstrated the applicability of our approach, specifically 
in the context of large-scale networks. Moreover, the results showed that the 
number of D2D users that can be served depends on QoS requirement of cellular 
users. If no QoS requirement exists, serving all D2D users causes degradation 
of the cellular aggregate utility, depending on the channel qualities as well 
as the number of D2D users. In addition, it was concluded that using Q-learning 
better-reply dynamics results in a fast convergence to equilibrium.

\section{Appendix}\label{sec:Appendix}
%
\subsection{Proof of Proposition \ref{pr:lowerBound}}\label{sec:AppOne}
According to our system model, $p_{k} \leq p_{d}^{(M)}~\forall~k \in \mathcal{K}$. Moreover, 
$h_{uv,q}=f_{uv,q}g_{uv}$ with $0<f_{uv,q} \leq 1$ and $0 < g_{uv} \leq 1$. 
Hence,
\begin{equation}
\label{eq:AppOneFirst}
\sum_{q=1}^{Q}\sum_{l \in \mathcal{L}_{q}} \log \left (\frac{p_{c}h_{bl,q}}{1+\sum_{k \in \mathcal{K}_{q}} p_{k} h_{kl,q}}  \right )\geq
\sum_{q=1}^{Q}\sum_{l \in \mathcal{L}_{q}} \log \left (\frac{p_{c}h_{bl,q}}{1+\sum_{k \in \mathcal{K}_{q}}p_{d}^{(M)} g_{kl}}\right ).
\end{equation}
By basic properties of the logarithm, the right-hand side of (\ref{eq:AppOneFirst}) 
can be written as 
\begin{equation}
\begin{aligned}
\sum_{q=1}^{Q}\sum_{l \in \mathcal{L}_{q}} \log \left (p_{c}h_{bl,q} \right )-\sum_{q=1}^{Q}
\sum_{l \in \mathcal{L}_{q}}&\log\left (1+\sum_{k \in \mathcal{K}_{q}}
p_{d}^{(M)}g_{kl}\right)> \\ 
\sum_{q=1}^{Q}\sum_{l \in \mathcal{L}_{q}} \log \left (p_{c}h_{bl,q} \right )-\sum_{q=1}^{Q}
\sum_{l \in \mathcal{L}_{q}}&\sum_{k \in \mathcal{K}_{q}}
p_{d}^{(M)}g_{kl},\\
\end{aligned}
\end{equation}
where the inequality follows from the standard logarithm inequality, $\frac{a}{1+a}\leq
\log(1+a)\leq a,~\forall a > -1$ \cite{Love80}.

\subsection{Proof of Proposition \ref{pr:nCellular}}\label{sec:AppThree}
We proceed by contraposition, i.e., we show that if $\left \{q \in \mathcal{Q}|L_{q} \neq 1 \right \}
\neq \varnothing $ then the partitioning is suboptimal.

Let $\mathcal{C}$ be the set of all possible Q-way partitioning forms of $L+K$ vertices of 
$G_{E}$. Assume that there exists some partitioning $c \in \mathcal{C}$, by which the 
graph is partitioned into $Q_{a}$ clusters with $L_{q}> 1$. As $L=Q$ (see Section 
\ref{subsubsec:Network}), there remain $Q_{b}=Q-Q_{a}$ clusters with $L_{q}=0$. In what 
follows, we show that partitioning $c$ is suboptimal, by constructing another partitioning 
whose cost is less than that of $c$.

Index $Q_{a}$ and $Q_{b}$ clusters of partitioning $c$ by $1,...,Q_{a}$ and $Q_{a}+1,...,
Q$, respectively. Moreover, let $T_{a}$ and $T_{b}$ correspondingly denote the aggregate 
sum weight of edges inside all clusters with and without cellular users. Thus we have
\begin{equation}
T_{a}=\sum_{q=1}^{Q_{a}}\sum_{l \in \mathcal{L}_{q}}\left (\sum_{j \in \mathcal{L}_{q},
j\neq l}w_{jl}+\sum_{k \in \mathcal{K}_{q}}w_{kl} \right),
\end{equation}
and $T_{b}=0$ by Definition \ref{de:EWIG}. Let $T_{c}$ denote the total cost of partitioning 
$c$. In order to establish that partitioning $c$ is suboptimal, we show that
\begin{equation}
T_{c}=T_{a}+T_{b}> \underset{\mathcal{C}}{\min}\sum_{q=1}^{Q_{a}}\sum_{l \in \mathcal{L}_{q}}
\left (\sum_{j \in \mathcal{L}_{q},j\neq l}w_{jl}+\sum_{k \in \mathcal{K}_{q}}w_{kl} \right ).
\end{equation}
To this end, we construct some partitioning $c'$ with $T_{c'}<T_{c}$. Assume that we 
change only one cluster of $c$, say cluster $r \in \left \{1,...,Q_{a}\right\}$ with 
$L_{q}>1$, by removing a cellular user $J \in \mathcal{L}_{r}$. Since all vertices 
must be included in the partitioning, $J$ is added in some cluster $r' \in \left \{1,
...,Q\right \}-\left \{r\right\}$. Therefore, one of the following holds:
\begin{itemize}
\item $r' \in \left \{1,...,Q_{a}\right \}-\left \{r\right \}$, or 
\item $r' \in \left \{Q_{a}+1,...,Q \right\}$. 
\end{itemize}
It is clear that the first case results in the original problem. Hence, we assume 
that the cellular user $J$ is included in $r' \in \left \{Q_{a}+1,...,Q \right\}$, 
and refer to the new partitioning by $c'$. Then we have
\begin{equation}
T_{c'}=T_{c}-\sum_{j \in \mathcal{L}_{r}}w_{jJ}-\sum_{k \in \mathcal{K}_{r}}w_{kJ}+
\sum_{k \in \mathcal{K}_{r'}}w_{kJ}.
\end{equation}
Since $0\leq w_{kJ} \leq p_{d}^{(M)}$, we have $0\leq\sum_{k\in \mathcal{K}_{x}}w_{kJ}
\leq Kp_{d}^{(M)}$, for any clusters $x$. Moreover, as $\mathcal{L}_{r}>1$ and $w_{jJ}=C$ 
for $j,J \in \mathcal{L}$, then $\sum_{j \in \mathcal{L}_{r}}w_{jJ} \geq C$ (see also 
Definition \ref{de:EWIG}). Hence the worst-case occurs when: i)$\sum_{k \in \mathcal{K}_{r}}
w_{kJ}=0$, which means that in cluster $r$, no D2D user causes interference to the cellular 
user $J$, ii) $\sum_{k \in \mathcal{K}_{r'}}w_{kJ}=Kp_{d}^{(M)}$, that is, cluster $r'$ 
includes all D2D users that cause the maximum interference to the cellular user $J$, and 
iii) $\sum_{j \in \mathcal{L}_{r}}w_{jJ}=C$, i.e., $L_{r}=2$. As a result, 
\begin{equation}
\label{eq:New}
T_{c'}\leq T_{c}-C+Kp_{d}^{(M)}<T_{c},
\end{equation}
as we assume $C> Kp_{d}^{(M)}$ by Definition \ref{de:EWIG}. Therefore by (\ref{eq:New}) 
partitioning $c$ is suboptimal, which is the contraposition and hence the proof is complete.

\subsection{Proof of Proposition \ref{pr:conversion}}\label{sec:AppFour}
By Proposition \ref{pr:nCellular}, any optimal partitioning of the estimated 
network graph $G_{E}$ includes exactly one cellular user in each cluster; therefore 
we can assume that $w_{ij}=0$, $\forall i,j \in \mathcal{L}$. Moreover, by Definition \ref{de:EWIG}, 
$w_{ij}=0~\forall i,j \in \mathcal{K}$. Therefore we define a complete bipartite 
graph $G$ with $\mathcal{V}_{1}=\mathcal{K}$ and $\mathcal{V}_{2}=\mathcal{L}$. 
The weight of the edge connecting $k \in \mathcal{K}$ and $l \in \mathcal{L}$ 
is equal to the corresponding edge in $G_{E}$, i.e., $w_{kl}$. We then augment 
$\mathcal{V}_{2}$ by $K$ times replicating each node $l \in \mathcal{L}$, 
resulting in a set $\mathcal{L}'=\underbrace{\mathcal{L} \cup \mathcal{L}...
\cup \mathcal{L}}_{\times K}$. Using this set, a bipartite graph $G'$ is constructed, 
where $\mathcal{V}_{1}=\mathcal{K}$ and $\mathcal{V}_{2}=\mathcal{L}'$. The 
weight of an edge connecting any pair $k \in \mathcal{K}$ to every copy $l'
\in \mathcal{L}'$ of some $l$ is $w_{kl'}=w_{kl}$. On graph $G'$, a bipartite 
minimum-weighted matching results in a $K\times (K \times L)$ assignment 
matrix $\mathbf{B}=[b_{kl'}]$, so that the sum  
\begin{equation}
\label{eq:rep}
\sum_{k\in \mathcal{K}}\sum_{l\in \mathcal{L}'}w_{kl'}b_{kl'}
\end{equation}
is minimized. For each $l$, let the set of its copies be denoted by $\mathcal{U}_{l}$. 
Moreover, the set of all users $k \in \mathcal{K}$ that are assigned to any copy of $l$ 
is denoted by $\mathcal{A}_{l}$. Thus (\ref{eq:rep}) can be reformulated as
\begin{equation}
\label{eq:repTwo}
\sum_{l=1}^{L}\sum_{j\in \mathcal{U}_{l}}\sum_{j'\in \mathcal{A}_{l}}b_{jl}w_{jj'}b_{j'l}, 
\end{equation}
which is identical to (\ref{eq:clustObjectF}). Hence the proposition follows.

\subsection{Proof of Theorem \ref{th:PG}}\label{sec:AppTwo}
%
\subsubsection{Some Auxiliary Definitions and Results}
\label{subsec:AuxiliaryApp}
The proof is based on some auxiliary definitions and results that are briefly stated 
in the following. \\

In what follows, $v$ stands for a function defined on a discrete set $\mathcal{X}\subseteq 
\mathbb{Z}^{I}$ where $\mathcal{X}=\prod_{i\in I}\mathbf{x}_{i}$, $\mathbf{x}_{i}=
\left\{x_{i}\in\mathbb{Z}:\underline{x}_{i}\leq x_{i}\leq \overline{x}_{i} \right\}
\subseteq \mathbb{Z}$, and $\underline{x}_{i},\overline{x}_{i} \in \mathbb{Z}$. Moreover, 
$\left\|\mathbf{x}\right\|=\sum_{i}\left|x_{i}\right|$ denotes the $l_{1}$-norm of 
a vector $\mathbf{x}\subseteq \mathbb{Z}^{I}$.
\begin{definition}[Larger Midpoint Property (LMP)]
\label{de:LMP}
We say that a function $v:\mathcal{X} \rightarrow \mathbb{R}$ satisfies the larger midpoint property 
(LMP) if, for any $\mathbf{x},\mathbf{y} \in \mathcal{X}$ with $\left\|\mathbf{x}-\mathbf{y}\right\|=2$, 
\begin{equation}
\label{eq:LMPOne}
\underset{\mathbf{z}\in \mathcal{X}:\left\|\mathbf{x}-\mathbf{z}\right\|=\left\|\mathbf{y}-\mathbf{z}
\right\|=1}{\textup{maximum}}~~f(\mathbf{z})\geq tf(\mathbf{x})+(1-t)f(\mathbf{y})~~\left (\exists ~t\in (0,1) \right),
\end{equation}
or
\begin{equation}
\label{eq:LMPTwo}
\underset{\mathbf{z}\in \mathcal{X}:\left\|\mathbf{x}-\mathbf{z}\right\|=\left\|\mathbf{y}-\mathbf{z}\right\|=1}
{\textup{maximum}}~~f(\mathbf{z})\left\{\begin{matrix}
 >\min\left \{ f(\mathbf{x}),f(\mathbf{y})\right\}&\textup{if}~~f(\mathbf{x})\neq f(\mathbf{y})\\ 
 \geq f(\mathbf{x})=f(\mathbf{y})&\textup{o.w.}
\end{matrix}\right..
\end{equation}
\end{definition}
\begin{definition}[Separable Concave Function]
\label{de:SCF}
A function $v:\mathcal{X} \rightarrow \mathbb{R}$ is separable concave if it can 
be written in the form $v(\mathbf{x})=\sum_{i\in I}v_{i}(x_{i})$, where $v_{i}(x_{i})\geq
\frac{v_{i}(x_{i}-1)+v_{i}(x_{i}+1)}{2}$ for all $x_{i}\neq \underline{x}_{i},\overline{x}_{i}$.
\end{definition}
\begin{lemma}[\cite{Ui08}]
\label{le:SCFLMP}
If $v:\mathcal{X} \rightarrow \mathbb{R}$ is a separable concave function, then (\ref{eq:LMPOne}) 
holds, and therefore $v$ satisfies the larger midpoint property.
\end{lemma}
\begin{proposition}[\cite{Ui08}]
\label{pr:LMPNash}
Let $\mathfrak{G}$ be an exact potential game with a potential function $v$ that 
satisfies the LMP property. Then $\mathbf{i} \in \mathcal{I}$ maximizes $v$ if 
and only if it is a Nash equilibrium. 
\end{proposition}
%
\subsubsection{Proof of Theorem \ref{th:PG}}
\label{subsec:PotentialApp}
The proof consists of two parts. First we show that the power allocation game 
defined in Definition \ref{de:CPAG} is an exact potential game by deriving a 
potential function. This will prove the first part of Theorem \ref{th:PG}. 
Afterwards we establish that the potential function satisfies the LMP property, 
and we characterize the set of Nash equilibria using Proposition \ref{pr:LMPNash}. 
This will prove the second part of the theorem.\\
\textbf{Part One}\\
%
By Definition \ref{de:PG}, we need to find a function $v:\mathcal{I} \rightarrow 
\mathbb{R}^{+}$ that satisfies (\ref{eq:PG}). With $R_{k}(\mathbf{i})$ given by 
(\ref{eq:DDThroughput}) we have
\begin{equation}
\label{eq:Difference}
R_{k}(p_{k},\mathbf{p}_{-k})-R_{k}(p'_{k},\mathbf{p}_{-k})= \log 
\left(\frac{p_{k}}{p'_{k}}\right)-c(p_{k}-p'_{k})
\end{equation}
Define 
\begin{equation}
\label{eq:potential}
v(\mathbf{p}_{d,q})= \sum_{k \in \mathcal{K}_{q}} \log(p_{k})-
\sum_{k\in \mathcal{K}_{q}}cp_{k}.
\end{equation}
Then By simple calculus it follows that 
\begin{equation}
\label{eq:potentialDif}
v(p_{k},\mathbf{p}_{-k})-v(p'_{k},\mathbf{p}_{-k})=
 \log\left(\frac{p_{k}}{p'_{k}}\right)-c(p_{k}-p'_{k}).
\end{equation}
Therefore, according to Definition \ref{de:PG} and by comparing (\ref{eq:potentialDif}) 
with (\ref{eq:Difference}), it can be concluded that the power allocation game is an 
exact potential game with potential function defined in (\ref{eq:potential}).\\
\textbf{Part Two}
\begin{lemma}
\label{le:PCGameSCF}
The potential function of the cluster power allocation game (given by (\ref{eq:potential})) 
is separable concave.
\end{lemma}
\begin{IEEEproof}
Clearly, the potential function can be written as $v(\mathbf{p}_{d,q})=
\sum_{k\in \mathcal{K}_{q}}v_{k}(p_{k})$ with 
\begin{equation}
v_{k}(p_{k})=\log(p_{k})-cp_{k}.
\end{equation}
Thus, by the assumption $p_{k}>1$ (see Section \ref{subsubsec:Network}), we have
\begin{equation}
\begin{aligned}
\frac{v_{k}(p_{k}+1)+v_{k}(p_{k}-1)}{2}&=\frac{\log(p_{k}^{2}-1)-2cp_{k}}{2}\\ 
&\leq \frac{\log(p_{k}^{2})-2cp_{k}}{2}\\ 
&=\log(p_{k})-cp_{k}.
\end{aligned}
\end{equation}
Therefore, by Definition \ref{de:SCF}, the function is separable concave.
\end{IEEEproof}
\begin{lemma}
\label{le:PCGameLMP}
The potential function of the cluster power allocation game (given by (\ref{eq:potential})) 
satisfies the larger midpoint property.
\end{lemma}
\begin{IEEEproof}
The proof directly follows from Lemma \ref{le:SCFLMP} and Lemma \ref{le:PCGameSCF}. 
\end{IEEEproof}
Therefore, since the potential function satisfies the LMP property, the second part 
of Theorem \ref{th:PG} follows directly from Proposition \ref{pr:LMPNash}.  
\subsection{Proof of Proposition \ref{pr:PoSBound}}\label{sec:AppFive}
By Definition \ref{def:PoS}, $1\leq \textup{PoS}$. Hence we only need to show that 
$\textup{PoS}\leq\frac{\log\left (p_{d}^{(M)}\right)}{\log \left(\gamma_{\min}\right)}$. 
To this end, we need the following theorem.
\begin{theorem}[\cite{Nisan07}]
\label{th:PoSBound}
Let $\mathfrak{G}=\left\{\mathcal{K},\mathcal{I},\left\{R_{k}\right\}_{k \in \mathcal{K}}\right\}$ 
be a potential game with some potential function $v\left(\mathbf{i}\right)$. Also, let $f(\mathbf{i})
=\sum_{k=1}^{K}R_{k}(\mathbf{i})$. Assume that for any joint action profile $\mathbf{i}$,  
\begin{equation}
\frac{1}{\alpha }f(\mathbf{i})\leq v \left ( \mathbf{i} \right )\leq \beta f\left (\mathbf{i} \right ),
\end{equation}
for some positive constants $\alpha$ and $\beta$. Then PoS is at most $\alpha\beta$.
\end{theorem}
For the cluster power allocation game, we have $\mathbf{i} := \mathbf{p}_{d,q}$, and $v\left(\mathbf{p}_{d,q}\right)$ 
is given by (\ref{eq:potentialFunc}). Also, by the definition of utility function given in (\ref{eq:DDThroughput}), 
we have $f(\mathbf{p}_{d,q})=\sum_{k\in \mathcal{K}_{q}} \log(\gamma_{i})-\sum_{k\in \mathcal{K}_{q}}cp_{k}$. Besides, 
as $0<f_{uv,q} \leq 1$ and $0<g_{uv} \leq 1$ (see Section \ref{subsubsec:Network}), at each trial, for any selected 
transmit power $p_{k} \in \mathcal{M}$ and any player $k \in \mathcal{K}$, we have $\gamma_{min}\leq \gamma_{k} \leq 
p_{k}$. Therefore, for any $\mathbf{p}_{d,q}$,
\begin{equation}
\label{eq:large}
\frac{v(\mathbf{p}_{d,q})}{f(\mathbf{p}_{d,q})}\geq 1.
\end{equation}
On the other hand,
\begin{equation}
\begin{aligned}
\frac{v(\mathbf{p}_{d,q})}{f(\mathbf{p}_{d,q})}&=\frac{\sum_{k \in \mathcal{K}_{q}} \log(p_{k})-
\sum_{k\in \mathcal{K}_{q}}cp_{k}}{\sum_{k \in \mathcal{K}_{q}} \log(\gamma_{k})-
\sum_{k\in \mathcal{K}_{q}}cp_{k}}  \\ 
&< \frac{\sum_{k \in \mathcal{K}_{q}} \log(p_{k})}{\sum_{k \in \mathcal{K}_{q}} \log(\gamma_{k})}
< \frac{\log\left (p_{d}^{(M)} \right )}{\log\left(\gamma_{\min}\right)},
\end{aligned}
\end{equation}
where the first inequality is concluded from (\ref{eq:large}). Thus, by Theorem \ref{th:PoSBound}, 
the result follows.

%
%

\bibliographystyle{IEEEbib}
\bibliography{Myreferences}
\end{document}